 \documentclass[aps,pra,epsf,superscriptaddress,amsmath,amssymb,amsfonts,showpacs,nofootinbib,longbibliography]{revtex4-2}
\usepackage{amsmath}
\usepackage{amssymb}
\usepackage{bm}
\usepackage{graphicx}
\usepackage{color}
\usepackage[OT1]{fontenc}
\usepackage{subfigure}
\usepackage{braket}
\usepackage{physics}
\usepackage{ascmac}
\usepackage{ulem}
\usepackage{float}
\usepackage[top=30truemm,bottom=30truemm,left=25truemm,right=25truemm]{geometry}
\usepackage{graphicx}
\usepackage{docmute}
\usepackage{mathtools}
\usepackage{CJKutf8}
\usepackage{bbold}
\usepackage[T1]{fontenc}
\usepackage[utf8]{inputenc}
\usepackage[colorlinks=true,
            linkcolor=red,
            urlcolor=blue,
            citecolor=blue]{hyperref}

\usepackage{soul}
\usepackage{xcolor}

\makeatletter

		\@addtoreset{table}{section}
\makeatother

\usepackage{afterpage}

\begin{document}

\makeatletter
\def\@maketitle{
  \newpage
  \null
  \vskip 1em
  \begin{center}%
    {\LARGE \@title \par}%
    \vskip 1.5em%
    {\large
     \lineskip .5em%
     \begin{tabular}[t]{c}%
       \@author
     \end{tabular}\par}%
    \vskip 1em%
    {\large \@date}%
  \end{center}%
  \vskip 0pt  
}
\makeatother

\title{Droplet-gas phases and their dynamical formation\\ in particle imbalanced mixtures}

\author{J. C. Pelayo
}
\thanks{corresponding author email: jose-pelayo@oist.jp}
\affiliation{Quantum Systems Unit \& Okinawa Center for Quantum Technologies, Okinawa Institute of Science and Technology Graduate University,
Okinawa, Japan 904-0495}

\author{G. A. Bougas}
\affiliation{Department of Physics and LAMOR, Missouri University of Science and Technology, Rolla, MO 65409, USA}

\author{T. Fogarty}
\affiliation{Quantum Systems Unit \& Okinawa Center for Quantum Technologies, Okinawa Institute of Science and Technology Graduate University,
Okinawa, Japan 904-0495}

\author{Th. Busch}
\affiliation{Quantum Systems Unit \& Okinawa Center for Quantum Technologies, Okinawa Institute of Science and Technology Graduate University,
Okinawa, Japan 904-0495}

\author{S. I. Mistakidis}
\affiliation{Department of Physics and LAMOR, Missouri University of Science and Technology, Rolla, MO 65409, USA}

\date{\today}

\begin{abstract}

We explore the ground state phase diagram and nonequilibrium dynamics of genuine two-component particle-imbalanced droplets in both  isotropic and anisotropic three-dimensional confinements. 
A gradual transition from mixed droplet-gas to gas configurations is revealed 
as the average intercomponent attraction decreases or the transverse confinement becomes tighter.
Within the mixed structures, a specific majority fragment binds to the minority droplet, satisfying the density ratio locking condition, while the remaining atoms are in a gas state.
Our extended Gross-Pitaevskii numerical results are corroborated by a suitable variational approximation capturing the shape and characteristics of droplet-gas fragments. 
The tunability of the relatively low gas fraction is showcased through parametric variations of the atom number, the intercomponent imbalance, the trap aspect ratio, or the radius of a box potential. 
To validate the existence and probe the properties of these exotic phases,  we simulate the standard time-of-flight and radio frequency experimental techniques. 
These allow to dynamically identify the resilience of the droplet fragment and the expansion of the gas fraction. 
Our results, amenable to current experimental cold atom settings, are expected to guide forthcoming investigations aiming to reveal unseen out-of-equilibrium droplet dynamics.

\end{abstract}

\maketitle

\section{Introduction} \label{sec:Intro}

Ultracold quantum gases have proven to be versatile platforms for controllably simulating phenomena arising in disparate disciplines, ranging from polarons in condensed matter~\cite{grusdt2024impurities,massignan2025polarons,mistakidis2022physics}, to  Hawking radiation in astrophysics~\cite{steinhauer2016observation,schutzhold2025ultra}.
The hydrodynamic formulation of quantum gases~\cite{tsatsos2016quantum} allows also the study of processes related to fluid mechanics such as turbulence~\cite{tsatsos2016quantum,madeira2020quantum}.
Recently, liquid type configurations known as quantum droplets~\cite{bottcher2020new,luo2021new,mistakidis2023few} were shown to exist in 
bosonic quantum gases. 
These are self-bound states that arise for attractive interactions
and exist solely in the presence of quantum fluctuations that are repulsive in three-dimensions (3D). These fluctuations can therefore compensate the wave collapse, and they are modeled by the first-order Lee-Huang-Yang (LHY) quantum correction term~\cite{petrov2015quantum,petrov2016ultradilute}. 
Hence, these liquid structures are promising candidates for probing unseen beyond mean-field processes but also surface tension and incompressibility.     
Experimentally, although these states suffer from un-avoidable lossy mechanisms~\cite{semeghini2018self,d2019observation,ferioli2019collisions,cavicchioli2025dynamical}, they have been detected first in long-range dipolar gases~\cite{ferrier2016observation,politi2022interspecies,chomaz2022dipolar} and later in bosonic mixtures with short-range interactions~\cite{semeghini2018self,cabrera2018quantum,d2019observation,cheiney2018bright,burchianti2020dual}. 

A standard assumption during the formation of droplets 
in short-range interacting gases, is their tendency to maintain a fixed density ratio which has been proven~\cite{petrov2015quantum} to favor their stability and act against inherent self-evaporation mechanisms. 
Accordingly, their properties, including for instance, quasi-elastic collisions~\cite{ferioli2019collisions,hu2022collisional,debnath2023interaction,katsimiga2023interactions}, coexistence with non-linear excitations~\cite{li2018two,yougurt2023vortex,edmonds2023dark,katsimiga2023solitary,Bougas_2Ddrops}, and collective modes~\cite{tylutki2020collective,astrakharchik2018dynamics} have been analyzed by retaining this restriction. 
The latter leads to an effective single-field extended Gross-Pitaevskii (eGPE) description that predicts a transition from a Gaussian type to a flat-top droplet distribution~\cite{astrakharchik2018dynamics,petrov2015quantum} for increasing atom number and fixed interactions. 
Studies going beyond this condition have only very recently been put forth
e.g. revealing mixed droplet-gas configurations in 3D~\cite{flynn2023quantum,flynn2024harmonically}, spectral stability in one-dimension~\cite{Charalampidis_2comp}, existence of higher-lying multipole droplet states~\cite{Kartashov_multipoles} and the formation of droplets atop a super Tonks--Girardeau gas confined in a lattice potential~\cite{valles2024quantum}. 

Focusing on 3D geometries, there are several key questions that remain unclear. 
Here, the Gaussian or flat-top configurations of the effective single-field scenario are replaced by mixed droplet-gas configurations which appear in the presence of intercomponent population imbalance~\cite{englezos2023particle,flynn2023quantum,flynn2024harmonically} or prominent intracomponent interaction imbalance~\cite{gangwar2025interaction}. 
In essence, the minority component is self-bound and remains within the majority component forming a droplet, while the remaining excess atoms reside in a gas state~\cite{flynn2023quantum}. 
However, our understanding of the characteristics and controllability of these states is far from complete. 
As such, it would be interesting to unveil, for instance, the role of the attractive mean-field interactions and the transverse confinement on the structure and nature of these configurations. The confinement may even effectively adjust the dimensionality from 3D to lower dimensions and therefore provide a knob to adjust the droplet and gas regions but also realize transitions of self-bound to trapped gas configurations similar to effects known in one-dimension~\cite{Charalampidis_2comp}. 
An additional fundamental question concerns the construction of the involved droplet and gas constituents from a variational ansatz (VA) to gain more insights into their properties. 
Furthermore, the dynamical response of these structures is almost completely unexplored with the exception of their collective breathing mode~\cite{flynn2023quantum,flynn2024harmonically}. 
Imbalanced configurations are naturally expected to reveal dynamical response regimes absent in their balanced counterparts. 

To tackle the above issues, we consider a two-component bosonic mixture featuring different particle imbalance ratios (violating the density locking restriction) and which are confined in either isotropic or anisotropic 3D external traps. 
Our model is described by a set of coupled eGPEs in 3D~\cite{petrov2015quantum}, and we explore parameter regimes that are accessible in recent droplet experiments~\cite{cabrera2018quantum,semeghini2018self}, in contrast to previous theoretical studies 
focusing on arbitrary interaction regimes, trap frequencies and species~\cite{flynn2023quantum,flynn2024harmonically}.
Specifically, we investigate the droplet-gas crossover using scattering lengths obtained from the Feshbach spectroscopy of these experiments. We also study the largely unexplored impact of anisotropic trapping potentials and varying population imbalances on the emergent droplet configurations.
Focusing on extreme particle imbalances, we reveal clear mixed droplet-gas phases for increasing average mean-field attraction. 
Here, only a specific number of majority atoms bind to the minority droplet core, while any excess particles end up in the gas fraction surrounding the majority droplet fragment. 
Surprisingly, the majority-minority droplet atom number ratio is found to follow the density ratio locking condition. 
For weaker attractive averaged mean-field interactions these mixed configurations gradually deform into the gas phase. 
We also show that anisotropic traps enforcing a pancake geometry favor a spatially extended gas fraction that attains higher densities compared to the isotropic scenario. 
The binding energy of these exotic configurations is stronger for either smaller trap aspect ratios or reduced intercomponent imbalance, where in the latter case only droplets build atop both components.   
Another main result of our study is the construction of a VA which adequately captures the shapes of the droplet and the gas fragments over the entire parametric region of their existence and also allows one to find analytical predictions for their behavior. 
It follows, for instance, that the gas fraction is
enhanced by employing smaller box traps or larger atom numbers, in accordance with the eGPE simulations.

To further understand the properties of these mixtures and explore their dynamical response we simulate their time-of-flight (TOF) expansion~\cite{cabrera2018quantum,cavicchioli2025dynamical} and investigate their dynamical formation by employing a radio frequency (RF) particle transfer   technique~\cite{semeghini2018self,cabrera2018quantum}.
The TOF process refers to the sudden release of the external confinement which enables a clear  differentiation of the individual constituents. 
We find that the droplet fragments of both components remain intact throughout the evolution confirming their self-bound character, while the gas region undergoes free expansion. 
The expansion boundaries of the gas fraction are nicely captured by 
the hydrodynamic prediction of a single-component quantum gas in the Thomas-Fermi limit~\cite{Dalfovo_theory_1999},
thus further affirming the gaseous nature. 
Finally, following a RF sudden transfer process  between the components, it is possible to monitor the dynamical formation of the mixed phases. 
Here, the gas region surrounding the droplet kernel of the majority component forms almost simultaneously with the droplet.

This work is organized as follows. Section~\ref{sec:Setup} introduces the properties of the two-component droplet setup and the set of coupled 3D eGPEs used to compute the ground state as well as the nonequilibrium droplet dynamics. 
In Section~\ref{sec:Ground} we analyze the resultant mixed droplet-gas configurations and their transition to a gaseous phase upon different parametric variations. 
The characteristics of the mixed droplet-gas states are further explored via a variational ansatz in Section~\ref{Sec:Variational}. 
The existence of the droplet and gas regions are confirmed through simulations of TOF expansion in Section~\ref{sec:Dynamics} along with the dynamical nucleation of these mixed self-bound states by simulating RF population transfer from one to two hyperfine states. 
We provide a summary of our results and discuss future reasearch directions in Section~\ref{sec:Conclusions}. 
Appendix~\ref{sec:three_body_loss} addresses the impact of three-body losses in the droplet dynamics induced by the RF protocol.
Appendix~\ref{sec:droplet_reconstruction} elaborates on the dynamical reconstruction of the mixed droplet-gas phase through a suitable three-component GPE-eGPE system.

\section{Two-component 3D droplet setting} \label{sec:Setup}

We consider a homonuclear $(m_1=m_2=m)$ two-component Bose-Bose mixture composed by the two hyperfine states  $\ket{1} \equiv \ket{F=1,m_F=-1}$ and $\ket{2} \equiv \ket{F=1,m_F=0}$ of $^{39}$K, utilized in the recent 3D droplet experiments of Refs.~\cite{cabrera2018quantum,semeghini2018self}. 
The system is confined by an external harmonic potential, typical in cold atom experiments,  characterized by frequencies $(\omega_x,\omega_y,\omega_z)=2 \pi \times (5,5,5)$Hz, unless stated otherwise e.g.  when the impact of different aspect ratios is explored [Section~\ref{sec:Ground}]. 
The mean-field interactions are parameterized by the effective coupling strengths $g_{ij} = 4\pi\hbar^2a_{ij}/m$ with $i,j=\{1,2\}$ labeling each component, and with $a_{ij}$ denoting the corresponding 3D s-wave scattering lengths that are experimentally tunable via Fano-Feshbach resonances~\cite{chin2010feshbach,semeghini2018self,cabrera2018quantum}. The scattering lengths $a_{11}$ and $a_{12}$ vary slowly with the Feshbach magnetic field as compared to $a_{22}$ in the range where two-component droplets form (see also the magnetic field dependence of $a_{ij}$ in \cite{cabrera2018quantum,semeghini2018self}).
Close to the $B \approx 57 ~ \rm{G}$ intraspecies Feshbach resonance of the $\ket{2}$ state, $\delta g = g_{12} + \sqrt{g_{11}g_{22}} < 0$, and the mean-field theory predicts a wave collapse~\cite{donley2001dynamics,roberts2001controlled}. 
This behavior is circumvented and the mixture is stabilized by the presence of repulsive quantum fluctuations, leading to the formation of Bose-Bose quantum droplets when the condition $0 < -\delta g \ll g_{ii}$ 
is fulfilled~\cite{petrov2015quantum,luo2021new}. The left part of the inequality ensures that the system lies slightly beyond the mean-field stability regime and the right inequality implies that the magnitude of the  intercomponent attraction is much smaller than the intracomponent repulsion. 
Even though this condition is sufficient, it is not always necessary for stable droplet formation, since the latter depends also on the particle number~\cite{petrov2015quantum,semeghini2018self}.

The quantum fluctuations 
are modeled by the first-order LHY correction term~\cite{petrov2015quantum}, 
and the energy functional of the Bose-Bose mixture is expressed 
as 
\begin{subequations}
\begin{align}
E &= \int d^3\boldsymbol{r}\left[\sum_{i=1,2}\left(\frac{\hbar^2}{2m}|\nabla\psi_i|^2 + \frac{m}{2}(\sum_{k=x,y,z} \omega_k^2 k^2)n_i\right)+\mathcal{E}_{MF}+\mathcal{E}_{LHY}\right] \label{eq:energy_functional}, \\
\mathcal{E}_{\text{MF}} &= \frac{1}{2}g_{11}n_1^2 + \frac{1}{2}g_{22}n_2^2 + g_{12}n_1n_2 \label{eq:MF_energy_density}, \\
\mathcal{E}_{\text{LHY}} &= \frac{8m^{3/2}}{15\pi^2\hbar^3}(g_{11}n_1)^{5/2}f\left(\frac{g_{12}^2}{g_{11}g_{22}},\frac{g_{22}n_2}{g_{11}n_1}\right), \label{eq:LHY_energy_density} \\
f(x,y) &=  \frac{1}{4\sqrt{2}} \sum_{\pm}\left(1+y\pm\sqrt{(1-y)^2+4xy}\right)^{5/2}. \label{eq:f_function}
\end{align}
\end{subequations}
\noindent
The macroscopic 3D wave function of the $i$-th component is represented by $\psi_i$ and hence the 
densities are given by $n_i = |\psi_i|^2$. Each $\psi_i$ is normalized to the particle number, that is, $N_i = \int d^3\boldsymbol{r}\,n_i$, with the total particle number being $N=N_1+N_2$. The first and second terms in the energy functional describe the kinetic and trapping potential contributions, while the third and fourth terms refer to the mean-field and LHY energies respectively.
Throughout this work, we assume that the mixture is close to the 
mean-field stability edge namely 
$g_{12}^2 \approx  g_{11}g_{22}$ in Eq.~\eqref{eq:LHY_energy_density} which is also the regime that the experimental works have been operating~\cite{semeghini2018self,cabrera2018quantum}. This assumption not only simplifies the LHY term, since $f(x,y) \approx (1 + y)^{5/2}$ in Eq.~\eqref{eq:f_function}, but also avoids the issue of a complex LHY contribution which would signal droplet instability~\cite{petrov2015quantum,ancilotto2018self,Cui_droplet_2021}. 

In the 
absence of an external trap and in the thermodynamic limit, the energy density of Eq.~(\ref{eq:energy_functional}) is minimized if the bound atoms of the mixture follow the density ratio~\cite{petrov2015quantum} $n_1/n_2 = \sqrt{g_{22}/g_{11}}$ 
associated with the saturation densities,

\begin{equation}
n_i^{(0)} = \frac{25\pi}{1024}\frac{(\delta a)^2}{a_{11}a_{22}\sqrt{a_{ii}}(\sqrt{a_{11}}+\sqrt{a_{22}})^5},
\label{Eq:Dens_thermodynamic}
\end{equation}
where 
$\delta a = \sqrt{a_{11}a_{22}} + a_{12}$ quantifies the %
averaged mean-field interaction which will be extensively used in what follows. 
For increasing total particle number, $N$, the flat-top densities of $n_i$ approach $n_i^{(0)}$ and thus the density ratio locking translates to $N_1/N_2 = \sqrt{g_{22}/g_{11}} = \sqrt{a_{22}/a_{11}}$. 
This condition defines the \textit{particle balanced} quantum droplet setup in this work~\cite{flynn2023quantum,flynn2024harmonically}. 
As the system deviates from  density ratio locking, it becomes  \textit{particle imbalanced}. 
In contrast to previous two-component droplet studies~\cite{flynn2023quantum,flynn2024harmonically} which introduce
particle imbalance 
by adding particles to one of the components of an otherwise balanced mixture, 
we fix the total atom number 
and instead vary the 
ratio 
$N_1/N_2 = \{0.6N/0.4N,0.7N/0.3N,0.8N/0.2N\}$ labeled, for convenience,  
as $\{60/40, 70/30, 80/20\}$. 
A particular imbalanced configuration can be 
realized in experiments by using the RF  population transfer technique~\cite{semeghini2018self,cabrera2018quantum} which we numerically simulate in Section~\ref{rf_droplets} or Raman population transfer schemes~\cite{bakkali2021realization}. 

To determine the ground state and dynamical properties of the mixture 
the energy functional of  Eq.~\eqref{eq:energy_functional} is minimized with respect to both field components leading to the coupled eGPEs

\begin{equation}
i\hbar\frac{\partial}{\partial t}\psi_i = \left(-\frac{\hbar^2 \nabla^2}{2m} +  \frac{m}{2}(\sum_{k=x,y,z} \omega_k^2 k^2) + g_{ii}n_i + g_{12} n_j  + \frac{4m^{3/2}}{3\pi^2\hbar^3 } g_{ii}(g_{11}n_1 + g_{22}n_2)^{3/2}\right)\psi_i. \label{eq:eGPE}
\end{equation}
\noindent
The first and second terms on the right hand side of Eq.~\eqref{eq:eGPE} are the kinetic and trapping potential contributions  respectively. 
The third and fourth terms describe the intra- and intercomponent mean-field density-density interactions, 
while the last term designates the LHY correction. Recall that in the particle balanced limit defined by the aforementioned density locking condition, these coupled equations reduce to an effective single component eGPE~\cite{petrov2015quantum,petrov2016ultradilute,Ma_quantum_2023,pelayo2025phases}. 
Within our numerical implementation, Eq.~\eqref{eq:eGPE} is solved in spherical (cylindrical) coordinates for the case of an isotropic (anisotropic along the $z$-direction) trap. 
A fourth-order Runge-Kutta method is used to numerically solve the coupled eGPEs in both imaginary and real time. 
The spatial discretization used is typically $dr = 0.016$ $l_r$  ensuring that the observables of interest remain unchanged to a good degree with increasing spatial resolution. Here, $l_r=\sqrt{\hbar/(m\omega_r)}$ denotes the harmonic oscillator length.

\section{Transition from mixed droplet-gas to gaseous phases} \label{sec:Ground}

The coexistence of gas and droplet components is a key feature in the particle imbalanced regime~\cite{flynn2024harmonically,flynn2023quantum,valles2024quantum}. 
A harmonic trap is employed in our system, that serves not only as a container for the gas fraction but also facilitates to study the impact of trap anisotropy in the imbalanced system. 
In particular, the trap is designed to be either 
isotropic ($\omega_r = \omega_z$) 
or anisotropic ($\omega_z / \omega_r >1$)  
about the radial plane, 
allowing to assess different mixed droplet-gas configurations.
The radial trapping frequency, $\omega_x = \omega_y \equiv \omega_r = 2\pi \times 5~ \rm{Hz}$, is chosen so that the planar  oscillator length $l_r$ is much larger than the droplet radius in the isotropic setup.

\subsection{Isotropic external confinement}\label{groundstate_isotropic}

\begin{figure}[tb]
\centering  
\includegraphics[width=1\linewidth]
{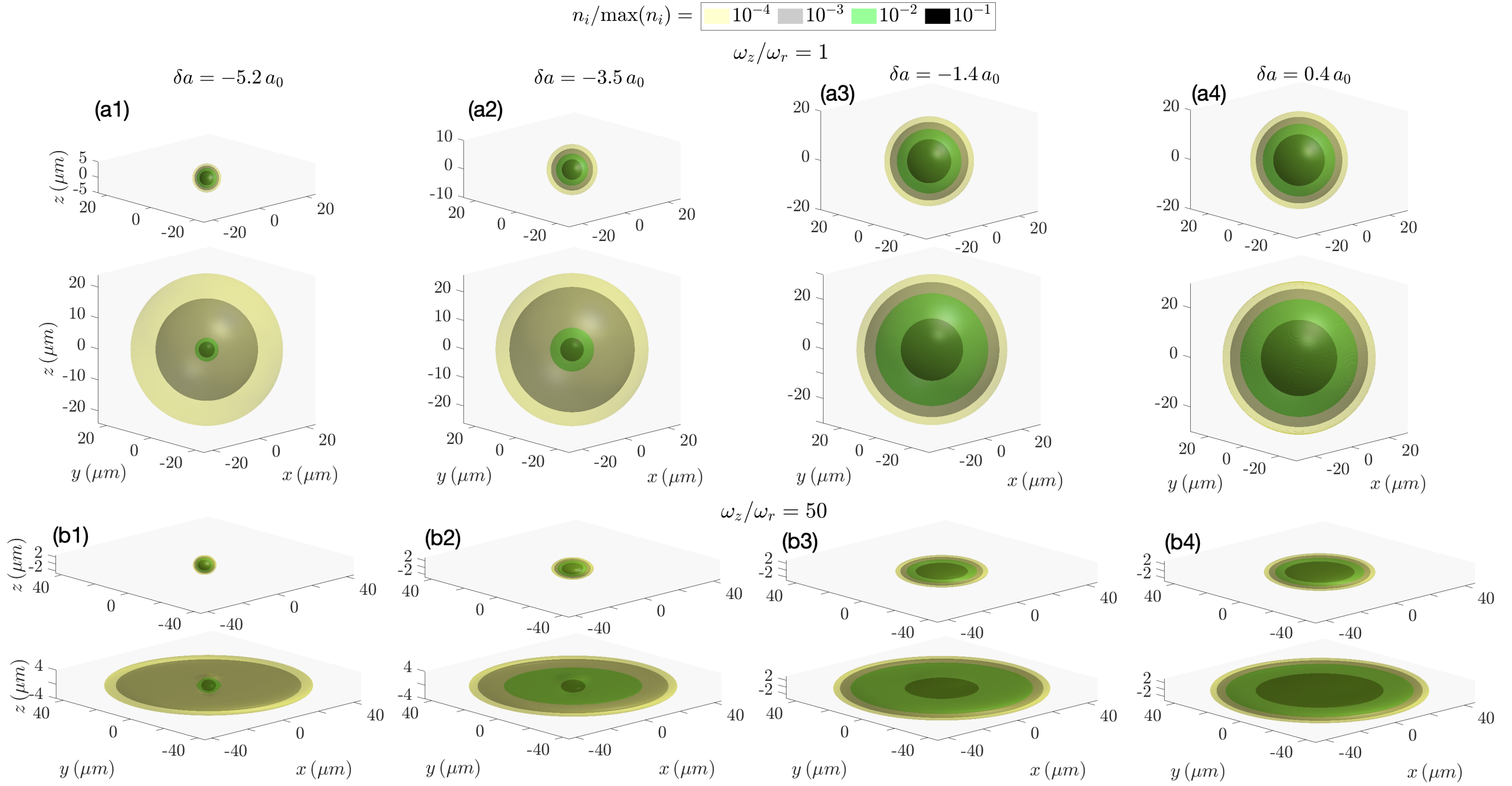}

\caption{Isosurfaces of the normalized 
densities of each component for the $N_1/N_2 = 80/20$ particle imbalanced Bose-Bose mixture 
with $N=2\times10^5$. 
Top panels (a1)-(a4) correspond to trap frequency aspect ratio $\omega_z/\omega_r = 1$ and bottom ones (b1)-(b4) refer to $\omega_z/\omega_r = 50$. 
Each column corresponds to different values of $\delta a$ [see labels at the top]. For every $[\omega_z/\omega_r, \delta a]$ combination the upper (lower) panels depict the normalized density of the minority (majority) component. 
Mixed droplet-gas configurations form upon  increasing the mean-field attraction (see panels (a1), (a2), (b1), (b2)) 
giving their place to pure gas states for larger $\delta a$ (panels (a3), (a4), (b3), (b4)).  
}
\label{fig:3D_densities}
\end{figure}

\begin{figure}[tb]
\centering  
\includegraphics[width=0.95\linewidth]
{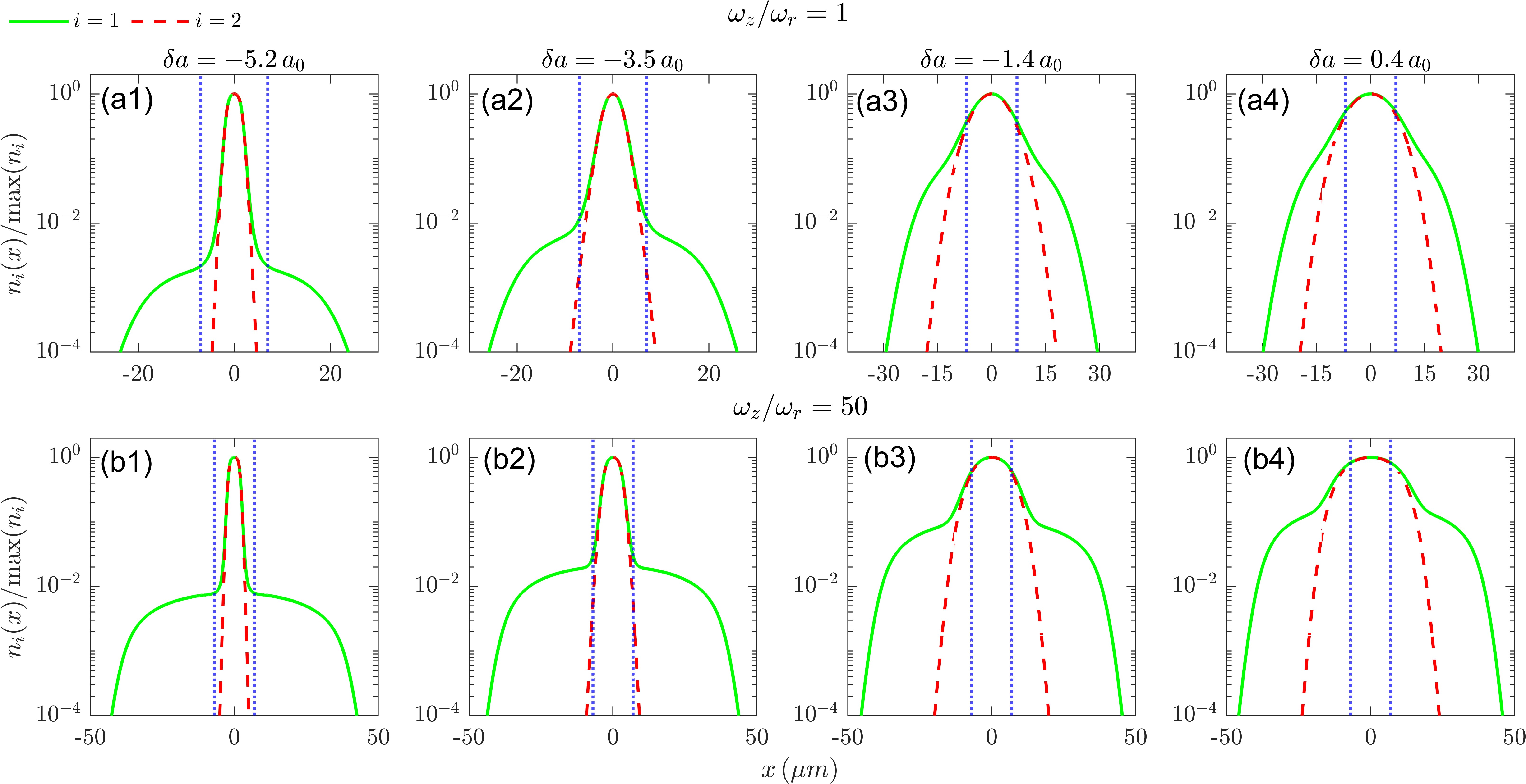}
\caption[]{
(a1)-(a4) [(b1)-(b4)] Normalized density profiles $n_i(x,0,0)/ \max (n_i(x,0,0))$ of each component for an isotropic  $\omega_z/\omega_r = 1$ [anisotropic, $\omega_z/\omega_r = 50$] 3D confinement. Each column refers to different values of $\delta a$ (see top row legends). 
The blue dotted lines indicate the edges of the planar harmonic oscillator length $\l_{r}=7\mu m$. 
The droplet core occurs always at distances smaller than $l_r$, while the gas portion extends beyond the oscillator length.
For anisotropic traps, the gas fraction extends only over the plane, and therefore gets more enhanced, compare upper and lower panels. 
A transition from a mixed droplet-gas to a gaseous phase takes place for less negative $\delta a$. 
In all cases, the particle imbalance is 80/20 with total atom number $N=2 \times 10^5$.  
 }
\label{fig:1D_density_slices}
\end{figure}

Focusing on the $80/20$ mixture, the  
different droplet phases can be discerned in the normalized density isosurfaces $n_i / \max  (n_i)$ presented in Fig.~\ref{fig:3D_densities} for various averaged mean-field interactions and trap aspect ratios. 
In the isotropic case $\omega_r = \omega_z$, and for large negative $\delta a$ [Fig.~\ref{fig:3D_densities}(a1), (a2)], both components form a pronounced droplet core (high density isosurface). 
Specifically, all atoms of the minority component [upper panels in Fig.~\ref{fig:3D_densities}(a1), (a2)] assemble in a self-bound droplet configuration within the majority species [lower panels in Fig.~\ref{fig:3D_densities}(a1), (a2)].
Interestingly, the concentric low-density isosurfaces surrounding the droplet kernel of the majority component refer to the extended gas fraction, thus manifesting the mixed droplet-gas state building on top of this component.
This arrangement is even more 
evident by inspecting the normalized density profiles along the $x$-direction, namely  $n_i(x,0,0)/ \max (n_i(x,0,0))$ [Fig.~\ref{fig:1D_density_slices}(a1), (a2)]. Here, the low-density gas fraction occurs only at positions larger than the oscillator length, $l_r \simeq 7~ \mu m$ [vertical dotted lines in Fig.~\ref{fig:1D_density_slices}].
However, the high-density droplet core of both components is confined at radii smaller than $l_r$ 
\footnote{Note that for larger particle numbers such as $N \sim 10^7$, the droplet cores may extend beyond the oscillator length, see e.g. Fig.~\ref{fig:GS_density}(a).}.
This provides a first evidence of the bound state characteristics of these configurations, 
see also our subsequent analysis for the ground state and Section~\ref{time_of_flight_dyn} for the dynamics.

\begin{figure}[tb]
\centering
\includegraphics[width=1\linewidth]
{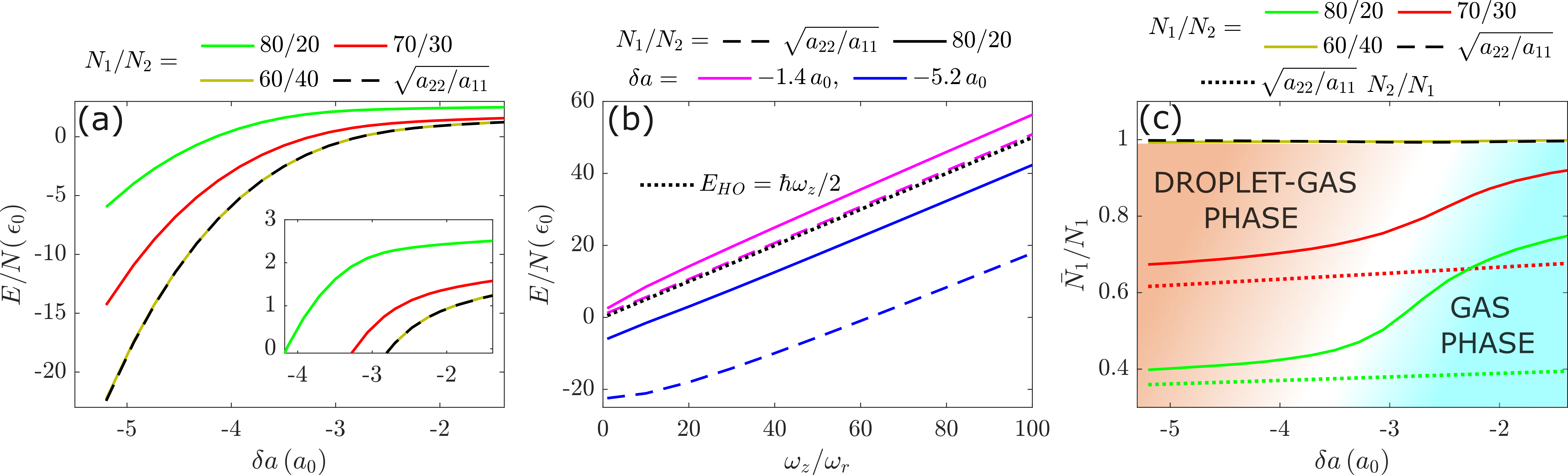}
\caption[]{
(a) Energy (total) per particle, $E/N$, as a function of $\delta a$ for different particle imbalances (see legend) and $\omega_z = \omega_r$. The inset focuses on $-4<\delta a<-1$ to highlight when the sign of $E/N$ changes from positive to negative. 
Stronger imbalances exhibit larger $E/N$ due to the presence of an enhanced fraction of excess particles. 
An excellent agreement occurs for $E/N$ between the system with 60/40 and the one obeying the density ratio locking (dashed lines).
(b) $E/N$ with respect to the trap anisotropy $\omega_z / \omega_r$ and different interaction strengths (see legend). The energy per particle increases with $\omega_z / \omega_r$, specifically as the 2D regime is  approached, due to the large potential energy in the $z$-direction quantified by $E_{HO}$ (dotted line) irrespective  of $N_1 / N_2$.
(c) Phase diagram of the droplet-gas and the pure gas phase in the plane ($\bar{N}_1/N_1 - \delta a$). The number of atoms populating the droplet fragment of the majority component, $\bar{N}_1 / N_1$, for varying $\delta a$ and different imbalances (see legend) is also depicted.  
A decreasing tendency is observed for larger imbalances. 
In all cases, an adequate estimate is given by employing the density ratio locking in the thermodynamic limit (dotted lines).
The comparison pertains only to the case where a droplet core is present, since for large $\delta a$ a pure gas phase occurs as indicated by the light blue shaded region. The phase boundary is determined by TOF discussed in Sec.~\ref{time_of_flight_dyn}.
}
\label{fig:energy_N_per_da}
\end{figure}

For larger $\delta a$ [Fig.~\ref{fig:3D_densities}(a3), (a4)], both components possess comparable density distributions,  
whose central part (previously signifying the droplet core) becomes significantly 
expanded.
In fact, the prominent central peak of the ensuing density profile of the majority component tends to disappear, 
see lower panels in Fig.~\ref{fig:3D_densities}(a3), (a4) and Fig.~\ref{fig:1D_density_slices}(a3), (a4).
Indeed, the extent of both components is rather dictated by the oscillator length, $l_r$,  implying that they display gas characteristics, see also the dynamical manifestation of this property via time-of-flight in Section~\ref{time_of_flight_dyn}. 
This behavior is also reflected in the energy per particle (in units of $\epsilon_0 \equiv \hbar \omega_r$) depicted in Fig.~\ref{fig:energy_N_per_da}(a), where for the $80/20$ mixture a larger $\delta a$ leads  to a positive energy [as also seen in the inset of Fig.~\ref{fig:energy_N_per_da}(a)].
In contrast, as $\delta a$ decreases, $E/N$ turns negative caused by 
the pronounced droplet fraction existing in both components\footnote{In general, for tighter transverse traps $|E/N|$ increases for stronger attractive $\delta a$. }.
Furthermore, in the case of increasing particle imbalance $N_1/N_2$, the energy per particle becomes overall less negative regardless of $\delta a$ [Fig.~\ref{fig:energy_N_per_da} (a)]~\cite{flynn2023quantum}.
As will be argued below, this is traced back to the fact that 
more and more excess particles in the majority component end up in the gas phase since less majority atoms can bind to the minority component, hence increasing the overall energy of the mixture. 
Note also that $E/N$ corresponding to the $60/40$ mixture matches almost exactly with the energy referring to the particle balanced droplet ($N_1/N_2 = \sqrt{a_{22}/a_{11}}$), due to the particular $a_{22} / a_{11}$ ratio in the considered $\delta a$ range\footnote{For different intervals of $\delta a$, i.e., $\delta a \approx -17\,a_0$, the energy per particle of the balanced droplet becomes close to the one of $N_1/N_2 =50/50$ configuration.}. As one increases the trapping frequency, the droplet core is largely unaffected for large negative $\delta a$ with its impact becoming  more prominent for the gas part and for increasing $\delta a$ (not shown here for brevity). The  impact of an external trap on the  quantum droplet configuration has  also been discussed in  Refs.~\cite{oldziejewski2020strongly,debnath2022dropleton}.

\subsection{Impact of trap anisotropy}\label{trap_anisotropy}

Next, we examine anisotropic two-component droplet configurations enforced by a finite trap aspect ratio between the planar and transverse $z$-direction namely 
$\omega_z > \omega_r$. 
In the case of large negative $\delta a$ [Fig.~\ref{fig:3D_densities}(b1), (b2)], an argument similar to the isotropic case can be made, 
meaning that the minority species atoms reside in the  droplet state [upper panels in  Fig.~\ref{fig:3D_densities}(b1), (b2)] and the majority component arranges into a prominent droplet fragment and a gas fraction [lower panels in Fig.~\ref{fig:3D_densities}(b1), (b2)]. 
However, the densities feature 
a pancake structure caused by 
the large anisotropy, $\omega_z / \omega_r = 50$. 
For instance, the droplet fragment in each component 
manifested by the high density isosurfaces in Fig.~\ref{fig:3D_densities}(b1), (b2) is 
squeezed along the $z$ direction due to the associated tight  trap frequency. 
In addition, it is still self-bound since its width is smaller than the planar oscillator length,  $l_r$, [see also the respective density profiles in Fig.~\ref{fig:1D_density_slices}(b1), (b2)].
The gas portion of the majority component, on the other hand, appears to be even more extended in space and attains higher density values when compared to the isotropic scenario [see in particular Figs.~\ref{fig:3D_densities}(a1), (a2) and~\ref{fig:1D_density_slices}(b1), (b2)].
Indeed, due to the trap anisotropy, the gas fraction is now allowed to extend 
almost exclusively along the 2D plane, restricted by $l_r$.
As a consequence, due to the presence of the trap together with the particle number conservation, the gas fragment reaches higher (than the $\omega_z / \omega_r = 1$ case) densities. 
Tuning the interactions towards the positive interaction regime, 
e.g. considering $\delta a = -1.4, ~ 0.4 ~ a_0$ presented in Fig.~\ref{fig:3D_densities}(b3), (b4) results in appreciable  larger widths 
of the minority component density.
Evidently, these are not droplet configurations, since their  width is comparable to $l_r$ [Fig.~\ref{fig:1D_density_slices}(b3), (b4)].
The corresponding majority component 
clearly features two separate structures, 
namely an inner high and an outer low density structure. 
These configurations are not droplets as can be seen from their energy per particle illustrated in Fig.~\ref{fig:energy_N_per_da}(b) but they are rather 
remnants of the droplet-gas coexistence existing for 
more attractive $\delta a$. 
Even though this behavior is also present 
in isotropic confinements  [Fig.~\ref{fig:1D_density_slices}(a3), (a4)], it is arguably more accentuated in the anisotropic case because the atoms are majorly distributed across the plane.

The self-bound character of the different states can be deduced by inspecting the energy per particle, $E/N$, shown in Fig.~\ref{fig:energy_N_per_da}(b) with respect to the trap aspect ratio. 
In this case, however, $E/N$ requires a different interpretation compared to
to the isotropic setting since it is naturally 
shifted towards positive values due to the large potential energy of the harmonic oscillator in the $z$-direction.
Therefore, even if $E/N$ at $N_1/N_2 = 80/20$, $\delta a = -5.2 ~ a_0$ and $\omega_z/ \omega_r = 50$ is positive, it still possesses a self-bound character because 
the energy per particle is smaller than the single particle energy $E_{HO} = \hbar \omega_z /2$ depicted by the dotted line in Fig.~\ref{fig:energy_N_per_da}(b). This is further supported by the corresponding TOF dynamics see, for instance, Fig.~\ref{fig:TOF_isotropic}(a3),(b3).
At $\delta a = -1.4 ~ a_0$ however, $E/N$ is above this threshold 
irrespective of the $N_1 / N_2$ ratio, further verifying that this is a purely gas state. 
Note that the energy corresponding to the $80/20$ configuration [dashed lines in Fig.~\ref{fig:energy_N_per_da}(b)] is always larger than the one assigned to the particle balanced droplet, due to the increased number of excess particles residing in the gas fraction of the majority component, see also Section~\ref{sec:core}.

\subsection{Estimation of the self-bound atom segment}  \label{sec:core}

A question that remains open regarding the droplet-gas phases identified above refers to which population 
of the majority atoms ends up in the gas and the droplet fractions. 
Since these two regions are complementary to each other it suffices to estimate one of them. For convenience, here, we compute the atom number in the droplet fragment by  
focusing on the isotropic external trap scenario and fixed $N=2 \times 10^5$. 
Recall also our discussion in Section~\ref{trap_anisotropy}, where it was found that 
trap anisotropy leads to an enhanced gas portion (due to the motion restricted along the 2D plane), meaning that anisotropy provides a knob to individually manipulate the gas and droplet segments, see more details on this topic in Sec.~\ref{box_pot}. 

It is also important to note at this point that in order to determine the boundaries of these fragments we need to rely on our eGPE simulations, for a specific total atom number, either by inspecting the structure of the ground state densities [see e.g. Fig.~\ref{fig:1D_density_slices}(a1)-(a4)] or monitoring their free expansion (see Sec.~\ref{time_of_flight_dyn}) since pure analytical arguments are not available.   
Indeed, according to our observations for $N=2 \times 10^5$, the gas fraction is approximately $0.1 \%$ of the maximum majority component density, see in particular Fig.~\ref{fig:1D_density_slices}(a1), (a2). 
For this reason we define the number of majority atoms bound in the droplet core as 
$\bar{N}_1 = 4\pi \int_0^{R_d} dr ~ r^2 n_1(r)$, where the radius\footnote{This criterion is similar to the one used in Ref.~\cite{flynn2024harmonically} for estimating 
the droplet fragment   
in each component.   
It depends on $N$ since enlarging the latter 
leads to enhanced gas densities and therefore different radii $R_d$.} $R_d$ corresponds to the width of the minority component which is set by the density at $n_2 / \max{(n_2)} \simeq 10^{-3}$.
Thus, the droplet core fraction refers to  $\bar{N}_1 / N_1$ and it is shown in  Fig.~\ref{fig:energy_N_per_da}(c) with respect to the average mean-field interaction. 
This measure is meaningful, of course, only within  the $\delta a$ regimes where droplet-gas coexistence occurs, e.g. Fig.~\ref{fig:1D_density_slices}(a1), (a2). In this sense, the light blue shaded region delineating the pure gas phase in Fig.~\ref{fig:energy_N_per_da}(c) is excluded.
The gas phase 
is strictly identified by the simulation of the experimentally relevant   
TOF process~\cite{cheiney2018bright,cavicchioli2025dynamical} allowing to classify droplet and gas states through their expansion dynamics,  
see for details Sec.~\ref{time_of_flight_dyn}.

As expected, for a particle balanced mixture  
all majority atoms reside  
in the droplet phase, i.e. $\bar{N}_1 / N_1 \approx 1 $, regardless of $\delta a$.
The same applies for the $60/40$ imbalance case, because it lies close to the $\sqrt{a_{22} / a_{11}}$ ratio within the studied $\delta a$ range, see the agreement between the black dashed and yellow solid lines in Fig.~\ref{fig:energy_N_per_da}(c).
However, for a larger $N_1 / N_2$  
the droplet fraction $\bar{N}_1 / N_1$ drops for constant $\delta a$, see Fig.~\ref{fig:energy_N_per_da}(c), reflecting the ever-increasing excess particle number accommodated in the gas portion of the majority component [see also Fig.~\ref{fig:3D_densities}(a1), (a2)]. 
To get an estimate for this behavior of the droplet core number, we rely on the observation that the droplet peak density of both components 
is approximately the same  
as $N_1/N_2$ increases for any fixed $\delta a$ (not shown for brevity).
In fact, their peak density values lie near the analytical predictions for the particle balanced system in the thermodynamic limit [Eq.~\eqref{Eq:Dens_thermodynamic}].
This implies that the ratio of the two density peaks satisfies to a certain extent 
the density locking condition, namely $n_1 / n_2 = \sqrt{a_{22} / a_{11}}$.
Given that the droplet core of both hyperfine levels roughly occupies the same volume [see also Fig.~\ref{fig:3D_densities} and~\ref{fig:1D_density_slices}(a1), (a2)], the self-bound fraction of the majority species is estimated to be $\bar{N}_1/N_1 = (N_2/N_1) \sqrt{a_{22} / a_{11}}$ [dotted lines in Fig.~\ref{fig:energy_N_per_da}(c)]. It is evident that this approximation indeed  
qualitatively captures the population $\bar{N}_1/ N_1$ for different particle imbalances.
Also, as $N$ increases 
the aforementioned agreement becomes gradually better, since the analytical estimate is based on the thermodynamic limit.
Finally, for fixed imbalance, the number of atoms occupying the droplet segment increases (decreases) for weaker (stronger) $\delta a$ attraction [Fig.~\ref{fig:energy_N_per_da}(c)]. 
This behavior is captured by the density locking condition, since $a_{22}$ reduces for more attractive averaged mean-field interactions, as was experimentally measured in Ref.~\cite{cabrera2018quantum}.

\section{Variational reconstruction of the mixed droplet-gas distributions}\label{Sec:Variational}

The existence of the above-discussed particle-imbalanced self-bound states containing droplet and gas segments has been thus far inferred from the numerical solutions of the two-component eGPEs of Eq.~(\ref{eq:eGPE}). 
It is, however, desirable to semi-analytically construct these mixed configurations in order to further elucidate their characteristics, such as the fact   
that the particle number in the majority droplet core follows approximately the density ratio locking, or obtain insights into the structure of the participating segments. 
To the best of our knowledge, such an approach remains still elusive in the current literature. 
Below, we develop an appropriate VA to analyze these phases for the isotropic scenario. 

\subsection{Variational ansatz and comparison to eGPE predictions}

The construction of the VA unfolds in two separate  stages. Namely, first one needs to capture the droplet core of the majority component and subsequently its 
gas fragment. 
To achieve this task, we initially relate the majority droplet core segment to the entire minority component distribution  through a variational parameter, $A$, satisfying  $\psi_1(\boldsymbol{r}) = A^{1/4} \psi_2(\boldsymbol{r})$.
The associated energy functional of Eq.~(\ref{eq:energy_functional}) can therefore be exclusively  expressed   
in terms of the minority species wave function.
To capture the involved droplet fragment, which is either gaussian or flat-top shaped~\cite{luo2021new}, we assume a super-Gaussian VA,  
\begin{equation}
\psi_2(\boldsymbol{r})  =   \frac{\sqrt{N_2} }{\sqrt{\frac{4}{3} \pi  \sigma_r^3 \Gamma \left(1+\frac{3}{2 m_r}\right)}}  e^{-\frac{1}{2} \left(\boldsymbol{r}/\sigma_r \right)^{2 m_r}}.
\label{Eq:Super_gaussian}
\end{equation}
Here, $m_r$ dictates the flatness of the droplet solution, i.e. in the case of $m_r=1$ it leads to a Gaussian which gradually transforms into a homogeneous configuration for $m_r \to \infty$, while $\sigma_r$ determines the width of the minority component distribution. 
This ansatz was argued to be adequate for describing the ground state properties of 1D~\cite{Otajonov_stationary_2019}, 2D~\cite{sturmer2021breathing,Otajonov_variational_2020} and quasi-2D droplets~\cite{pelayo2025phases}. 
Plugging the above expression into the energy functional expressed in terms of the minority wave function, we arrive to an 
energy functional depending on the following three variational parameters,
\begin{gather}
E[m_r,\sigma_r,A]  = \frac{\hbar^2}{m} \frac{\left(\sqrt{A}+1\right) \left(m_r^2 N_2 \Gamma \left(2+\frac{1}{2 m_r}\right)\right)}{2 \sigma_r^2 \Gamma \left(\frac{3}{2 m}\right)}+\frac{3 \left( \frac{g_{22}}{2} + \frac{g_{11}}{2} A  +g_{12} \sqrt{A}  \right) 2^{-\frac{3}{2 m_r}-2} N_2^2}{\pi  \sigma_r^3 \Gamma \left(1+\frac{3}{2 m_r}\right)}  \nonumber \\
+  m \omega_r^2  \left[  1+  \sqrt{A}  \right] \frac{3 N_2  \sigma_r^2 \Gamma \left(1+\frac{5}{2 m_r}\right)}{10 \Gamma \left(1+\frac{3}{2 m_r}\right)}
+ \frac{\hbar^2}{m} \frac{\sqrt{3} \left( g_{22}  +g_{11}  \sqrt{A}  \right)^{5/2} 2^{\frac{3}{2 m_r}+5} 5^{-\frac{3}{2 m_r}-1} N_2^{5/2}}{\pi  \sigma_r^{9/2} \Gamma \left(1+\frac{3}{2 m_r}\right)^{3/2}}.
\label{Eq:Energy_analytical}
\end{gather}
These variational  
parameters can be determined by requiring that the energy functional of Eq.~(\ref{Eq:Energy_analytical}) is minimized with respect to all of them simultaneously, i.e. $\frac{\partial E}{\partial m_r} = 0$,  $\frac{\partial E}{ \partial \sigma_r  } =0 $ and $\frac{\partial E}{\partial A}=0$. This process is followed below for extracting the droplet densities of the minority component, and thus also the droplet segment of the majority species for each combination of atom numbers and interactions.

To describe the gas segment of the majority component, we rely on the fact that it 
extends beyond the radial width of the minority component, see also Fig.~\ref{fig:1D_density_slices}(a1), (a2).
Therefore, at locations $\abs{\boldsymbol{r}} > \sigma_r$ the two components approximately decouple, and the gas wave function 
satisfies the standard time-independent  mean-field radial GPE
\begin{equation}
\mu_1 \varphi_1(\boldsymbol{r}) = -\frac{\hbar^2}{2m} \nabla^2 \varphi_1(\boldsymbol{r}) +\frac{1}{2} m \omega_r^2 \boldsymbol{r}^2 \varphi_1(\boldsymbol{r}) + g_{11} \abs{\varphi_1(\boldsymbol{r})}^2 \varphi_1(\boldsymbol{r}), \quad \abs{\boldsymbol{r}} > \sigma_r,
\label{Eq:Simplified_majority}
\end{equation}
where $\mu_1$ represents the chemical potential 
of the gas fraction building atop the majority component\footnote{The LHY correction has been omitted here since it is naturally assumed to be  negligible for a single-component weakly interacting dilute gas~\cite{Lopes_quantum_2017}. 
However, the LHY term is always taken into account when determining the droplet core [see also Eq.~\eqref{Eq:Energy_analytical}].
}.
Since by construction of the VA the number of particles accommodated in the majority droplet core is given by $N_2 \sqrt{A}$, then $\varphi_1$ is normalized to the excess particle number, i.e. $4\pi \int_{\sigma_r}^{\infty} dr ~ r^2 \abs{\varphi_1(\boldsymbol{r})}^2 = N_1 -N_2 \sqrt{A}$.
Given that the minimization process results in $A \approx 1.44$, the excess particle number becomes 
a significant fraction of the total particle number, 
especially for large imbalances.
Consequently, the Thomas-Fermi approximation is employed for $\varphi_1(\boldsymbol{r})$,  
leading to $\abs{\varphi_1(\boldsymbol{r})}^2 = \frac{m \omega_r^2 (R_{TF}^2-\boldsymbol{r}^2)}{2g_{11}} \theta( R_{TF}  - \abs{\boldsymbol{r}} )  $, where $\theta(\cdot)$ is the heaviside theta function and the Thomas-Fermi radius $R_{TF}$ being determined by the normalization condition.
As a last step, to reconstruct 
both fractions of the majority component, we propose the following interpolating wave function,
\begin{equation}
\phi_1(\boldsymbol{r}) = \sqrt{  \sqrt{A} \abs{\psi_2(\boldsymbol{r})}^2  + \abs{\varphi_1(\boldsymbol{r})}^2  \theta(  \abs{\boldsymbol{r}}  -\sigma_r ) },
\label{Eq:Majority_renormalized}
\end{equation}
which is normalized to $N_1$.

\begin{figure}[tb]
\centering  
\includegraphics[width=1\linewidth]
{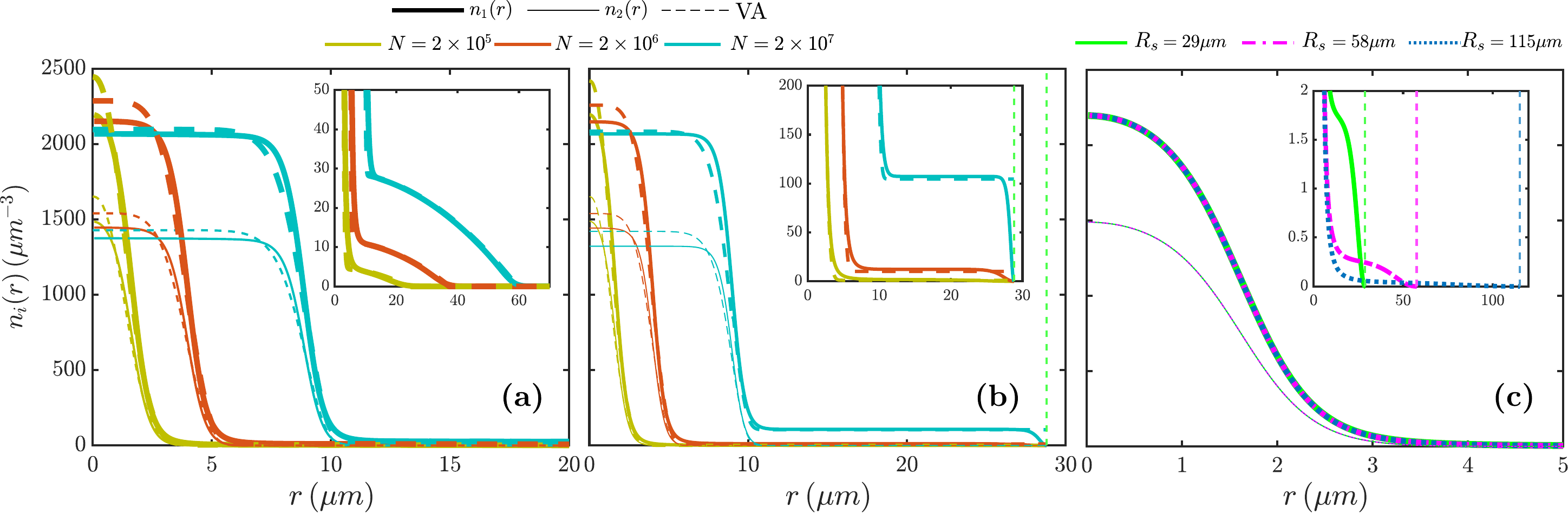}

\caption[]{
Comparison of the ground state two-component droplet densities (see legend) of a particle imbalanced setting computed via the VA (dashed lines) and the set of eGPEs (solid lines). 
The system features averaged mean-field interactions, $\delta a = -5.2 ~ a_0$, and fixed  imbalance, $N_1/N_2 = 80/20$, while the total particle number, $N$, increases (see legend) and the external confinement is an (a) isotropic spherical trap and (b) a hard sphere. 
The insets in panels (a), (b) provide a magnification of the majority component gas fraction. 
The agreement between the VA and the eGPE results for both the droplet and the gas fragments gradually improves for larger atom numbers.
By increasing the hard sphere radius, $R_s$, while holding  $N=2 \times 10^5$ constant in panel (c), the majority droplet core remains unchanged, while the gas segment (see in particular the inset) expands occupying the available space all the way towards the boundaries. These results are obtained within the eGPEs but a similar behavior is predicted with the VA (not shown for better visualization).
In panels (b) and (c) the vertical dashed lines mark the location of the radius of each hard sphere.
}
\label{fig:GS_density}
\end{figure}

A direct comparison between the VA outcome (dashed lines in Fig.~\ref{fig:GS_density}(a)) and the eGPE results (solid lines in Fig.~\ref{fig:GS_density}(a)) for the ground state densities of the isotropically confined two-component setting 
at interactions $\delta a = -5.2 ~ a_0$ and for imbalance $N_1/N_2 = 80/20$ is shown in Fig.~\ref{fig:GS_density}(a).
It can be readily seen that for increasing atom number, the agreement among the two methods improves significantly, 
especially for the majority component.
This behavior is expected since for larger $N$ 
we gradually approach the thermodynamic limit, where the density ratio locking condition employed for the construction of the majority droplet fragment becomes more adequate~\cite{petrov2015quantum}.  
Indeed, both components tend to acquire 
their corresponding saturated flat-top density profiles as $N$ increases, which are approximately given by the analytical formula of  Eq.~\eqref{Eq:Dens_thermodynamic}.

On the other hand, the gas fraction [magnified in the inset of Fig.~\ref{fig:GS_density}(a)] of the majority component is heavily dependent on the particle number, as predicted also from the VA, see also Eq.~(\ref{Eq:Majority_renormalized}).
Specifically, the Thomas-Fermi radius increases with $N$, a behavior encountered in single-component gases as well.
However, despite the large number of excess particles [see Fig.~\ref{fig:energy_N_per_da}(c)], the density of the gas fraction remains significantly lower than that of the droplet segment [see also Fig.~\ref{fig:1D_density_slices}(a1), (a2)].
This is because the gas fraction essentially extends over 
most of the available space of the spherical trap, thus lowering its density.
Finally, let us note that the comparison between the VA and the respective coupled eGPE results 
turns out to be very good also for different $N_1/N_2$ ratios, as well as different averaged mean-field interactions (not shown for brevity). The above instructive variational reconstruction of the majority component distribution (Eq.~\eqref{Eq:Majority_renormalized}), suggests  the modeling of the complex droplet-gas configurations by a coupled three-component GPE-eGPE system. This is described in   
~\ref{sec:droplet_reconstruction} and shown to accurately capture the dynamical formation of both droplet and gas segments.

\subsection{Droplets trapped in hard spheres}\label{box_pot}

Since the gas fraction adjusts to the shape of its external potential, it is reasonable 
to enhance its density 
by utilizing 
a small hard sphere external potential of radius $R_s$, see Fig.~\ref{fig:GS_density}(b). 
Experimentally, box type traps of arbitrary shape can be created using digital micromirror devices~\cite{SaintJalm,navon2021quantum,banerjee2024collapse}. 
It is evident that while the minority and majority droplet core profiles remain approximately the same with the isotropic trap setting, the gas fraction is enhanced, especially for larger atom numbers such as $N=2 \times 10^7$ [compare the insets of Fig.~\ref{fig:GS_density}(a) and (b)].
A prediction for the value of the almost flat density profile of the gas fragment can be made by invoking the decoupled gas equation [Eq.~\eqref{Eq:Simplified_majority}].
From the normalization of the wave function assigned to the excess particles one obtains $\abs{\varphi_1(\boldsymbol{r})}^2 = 3 \frac{N_1 -N_2 \sqrt{A}}{4\pi(R_s^3-\sigma_r^3)} $.
This profile is incorporated in the interpolated wave function of the majority component in the VA [Eq.~\eqref{Eq:Majority_renormalized}], which yields very good agreement with the results stemming from the coupled eGPEs [see the inset in Fig.~\ref{fig:GS_density}(b)]. 
Moreover, it can be seen that for fixed $R_s$ and larger $N$ the gas fraction becomes more pronounced, while the width (height) of the droplet core increases (slightly reduces), see Fig.~\ref{fig:GS_density}(b).
Eventually, when approaching the thermodynamic limit the droplet fragment reaches its saturation density and solely expands, accompanied by an increase of the gas fraction.

By the same token, the aforementioned analytic estimate for the $\abs{\varphi_1}^2$ density  profile implies that the gas fraction becomes 
suppressed in the case of an increasing 
radii of the hard sphere. 
This is explicitly verified within our eGPE simulations for larger $R_s$ and all other parameters kept fixed as shown in Fig.~\ref{fig:GS_density}(c) and in the corresponding inset depicting explicitly the spatial regime of the gas segment for better visualization.
Interestingly, the majority droplet fragment  remains almost unaffected by the sphere radii variations, and the same holds true for the minority component.

\section{Dynamical formation of mixed droplet-gas configurations} \label{sec:Dynamics}

Having explored the ground state properties of the mixed droplet-gas configurations, we now undertake two first steps to assess their dynamical formation. 
Specifically, we simulate the standard experimental techniques of TOF expansion dynamics and RF transfer. 
The former allows to further attest the existence of the droplet and gas segments as well as unveil the characteristics of their  dynamical response. 
On the other hand, the latter method permits to realize the dynamical formation of the aforementioned exotic configurations, and in particular how fast the droplet core develops and whether the gas fraction expands and accumulates.

\subsection{Time-of-flight expansion  dynamics: Droplet versus gas fragments}\label{time_of_flight_dyn}

\begin{figure}[tb]
\centering  
\includegraphics[width=1\linewidth]
{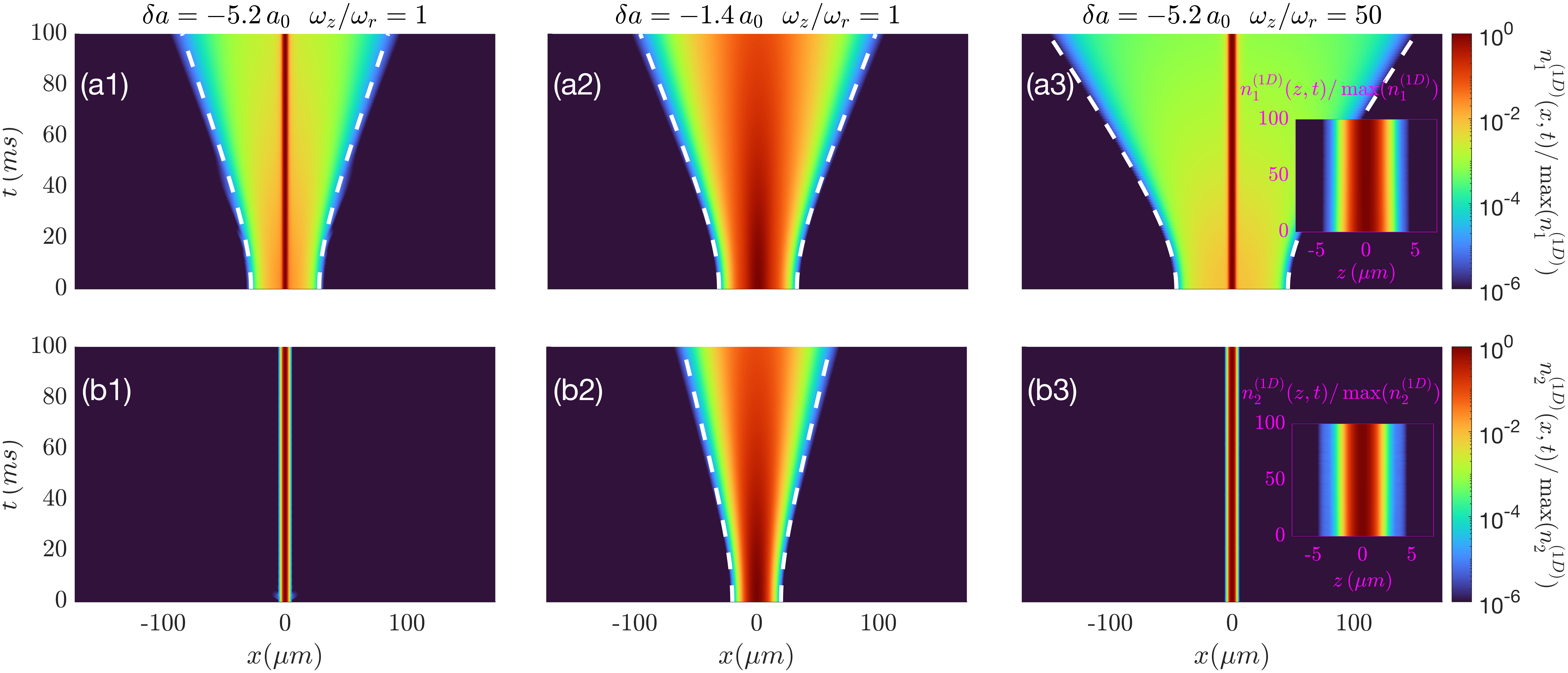}

\caption[]{Simulated TOF dynamics of the (a1)-(a3) majority and (b1)-(b3) minority component in the $N_1/N_2 = 80/20$ imbalanced configuration for different trap aspect ratios, $\omega_z / \omega_r$, and averaged mean-field interactions, $\delta a$, see legends. 
In all cases, the dynamics of the density profiles $n^{(1D)}_i(x)$ with $i=1,2$ (see main text) is presented.  
In the droplet-gas configuration [panels (a1), (b1)] the droplet segment quantified by the high-density central region  remains self-bound in the course of the evolution, while the surrounding gas of the majority component expands [see e.g. panel (a1)]. In sharp contrast, within the gas phase [panels (a2), (b2)], both components expand after the trap release. 
For the anisotropically trapped droplet-gas state [panels (a3), (b3)] the induced evolution follows a similar trend to the isotropic setting [panels (a1), (b1)].  
The density along the $z$ direction, $n^{(1D)}_i(z)$, remains intact as can be seen in the insets of (a3) and (b3). 
In all cases, the white dashed lines indicate the semi-analytical estimate of the expansion radius 
of a single-component Bose gas. 
}
\label{fig:TOF_isotropic}
\end{figure}

The most appropriate dynamical protocol for distinguishing a gas from a droplet state corresponds to the sudden release of the external trap in order to observe the free evolution of the atoms commonly referred to as TOF dynamics~\cite{bloch2008many,langen2015ultracold}. 
This method has been extensively used experimentally to assess the bound state characteristics of droplets~\cite{cabrera2018quantum,semeghini2018self}, but also to extract information pertaining to quantum gas phases, such as their momentum distributions~\cite{Davis_Bose_1995,Anderson_observation_1995}. 
Here, in order to understand the main dynamical properties of the mixed phases we focus on the $80/20$ particle imbalance setup which contains well-separated droplet and gas fragments. Other settings characterized by a smaller  imbalance exhibit similar responses but their gas fraction is less pronounced (not shown).

For the 3D isotropic case, the harmonic trap is released simultaneously (at $t=0$) in all three spatial directions and the ensuing two-component state is left to freely evolve, see Fig.~\ref{fig:TOF_isotropic}(a1)-(b2).  
At relatively strong averaged mean-field attractions, e.g. $\delta a = -5.2 ~ a_0$ shown in Fig.~\ref{fig:TOF_isotropic}(a1), (b1), the droplet fraction of both components (corresponding to the high-density central peak) remains unaltered throughout the TOF evolution with the mean-field energy having the dominant contribution.
This behavior manifests  
the inherent bound state nature of the droplet, as evidenced by the density profiles $n^{(1D)}_i(x,t)=n_i(x,0,0,t)$. 
Interestingly, the droplet core fraction of the majority component  $\bar{N}_1/ N_1$ is roughly the same with the one of the corresponding ground state configuration [see also Fig.~\ref{fig:energy_N_per_da}(c)] throughout the time evolution. 
In contrast, the gas segment of the majority component naturally expands in all spatial directions, see Fig.~\ref{fig:TOF_isotropic}(a1).
Recall that the shape of the gas fraction can be successfully modeled by a single component Bose gas [see also Sec.~\ref{Sec:Variational}]. 
For this reason, we are able to also put forward a semi-analytical estimate for the radius of the expansion dynamics. 
Based on the hydrodynamic formulation of the single component GPE, the time-dependent radius, $R(t)$, is determined by $R(t) = R(0)b(t)$, where $R(0)$ is the initial (at $t=0$) radius, and $b(t)$ called the scaling parameter satisfies the differential equation $\ddot{b}(t) = \omega_r^2 b^{-4}(t)$, with the obvious initial conditions $b(0)=1$, $\dot{b}(0)=0$~\cite{Dalfovo_theory_1999}.
The resulting dynamical behavior of the gas segment is indicated by the white dashed lines in Fig.~\ref{fig:TOF_isotropic}(a1), capturing to an excellent degree the expansion boundaries of the gas fraction.

Increasing the averaged mean-field interactions to $\delta a = -1.4 ~ a_0$, the two-component setting resides in a pure gas phase as argued at the ground state level in Sec.~\ref{groundstate_isotropic}, \ref{sec:core} and as hinted from the energy arguments depicted in Fig.~\ref{fig:energy_N_per_da}(a), (c).
As such, the release of the trap leads to a clean overall expansion in both components  [Fig.~\ref{fig:TOF_isotropic}(a2), (b2)].
Even though the semi-analytical estimate for the radius of the accompanied expansion relies on a single component GPE, it nicely captures the boundaries of the TOF dynamics of both components marked by the dashed lines in Fig.~\ref{fig:TOF_isotropic}(a2), (b2). 
Note that around $\delta a = -1.4~a_0$ the energy per particle of the ground state configuration is large and positive [inset of Fig.~\ref{fig:energy_N_per_da}(a)]. 
This is related to the fact that the attractive intercomponent interaction energy plays a minor role when compared to the rest  energy contributions being positive. 
Indeed, after the trap release these positive energy terms 
are predominantly  converted to kinetic energy which 
subsequently dictates  the expansion dynamics of both components.
Therefore, the TOF dynamics of the individual  components can be considered in this case as approximately decoupled.

Turning to the anisotropic case where $\omega_z / \omega_r = 50$, we apply a trap release solely along the planar direction, while maintaining the tight confinement in the $z$-direction. This is because we aim to describe the expansion of the in-plane 2D droplet and avoid more complicated effects arising from the release of the transverse direction leading to significantly faster dynamics along the $z$-direction and interfering with the in-plane evolution.
Below, we focus on the $80/20$ imbalanced setup characterized by $\delta a = -5.2 ~ a_0$ since it represents a clear droplet-gas state. 
To track the emergent planar and transverse dynamics, we employ the associated density profiles $n^{(1D)}_i(x,t)$ 
and $n^{(1D)}_i(z,t)=n_i(0,0,z,t)$ 
with  $i=1,2$, respectively. 
As expected, the dynamics is frozen along the $z$-direction due to the tight trap confinement as can be seen from the respective 1D integrated density profiles, $n^{(1D)}_i(z,t)$, shown in the insets of Fig.~\ref{fig:TOF_isotropic}(a3), (b3). 
This is, of course, not the case for the in-plane evolution. 
Similarly to the isotropic setting, the involved  droplet fraction of both components remains robust following the trap release [see the high density peaks of $n^{(1D)}_i(x,t)$ in Fig.~\ref{fig:TOF_isotropic}(a3), (b3)].
A prominent difference though occurs for 
the gas segment dynamics  
which appears to expand much faster from the one of the isotropic scenario, compare Fig.~\ref{fig:TOF_isotropic} (a1) and (a3).
Indeed, due to the larger radius of the gas segment in the anisotropic setting [see also Fig.~\ref{fig:1D_density_slices}(b1)], the originally enhanced   
potential energy is converted to a large kinetic energy contribution, thus leading to the fast expansion dynamics.
The associated expansion radius of the gas, $R(t)$, can again be adequately captured by the amended  hydrodynamic prediction 
$R(t) = R(0) b(t)$, where $\ddot{b}(t)=\omega_r^2 b^{-3}(t)$
and $b(0)=1$, $\dot{b}(0)=0$ \cite{Dalfovo_theory_1999,muga2009frictionless}. The resulting white dashed lines in Fig.~\ref{fig:TOF_isotropic}(a3) display a very good agreement with the instantaneous expansion boundaries stemming from the coupled eGPEs.

\subsection{Dynamical creation of the two-component droplet states via RF transfer}\label{rf_droplets}

\begin{figure}[tb]
\centering  
\includegraphics[width=1\linewidth]
{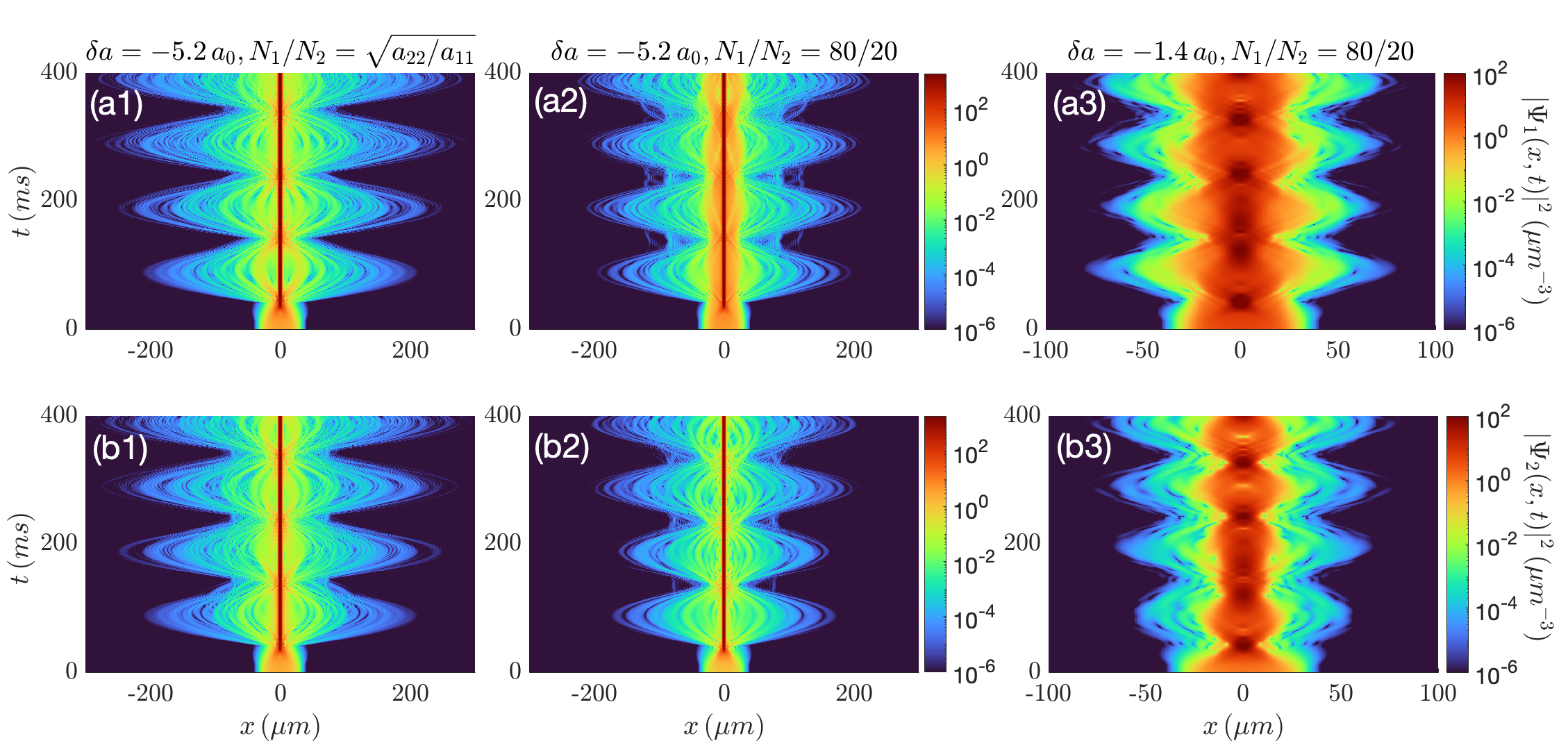}

\caption[]{Density dynamics following a RF atom transfer from a single-component gas to a two-component particle (a1)-(b1) balanced and (a2)-(b3) imbalanced mixture with $N_1/N_2 = 80/20$ and different mean-field interactions (see legends). Top (bottom) panels stand for the density evolution of the majority (minority) component.
For sufficiently large attractions and imbalanced settings [panels (a2), (b2)], a central excited droplet fragment is dynamically nucleated expelling particles due to self-evaporation and being surrounded by a gas segment. 
The latter is suppressed for particle balanced settings [panels (a1), (b1)]. 
In contrast, a collective excitation process occurs for smaller attractions [panels (a3), (b3)] where the system is genuinely gaseous. 
The total atom number is $N=2 \times  10^5$ and the external 3D trap has $(\omega_x, \omega_y, \omega_z)= 2 \pi \times (5,5,5) {\rm Hz} $.}
\label{fig:RF_transfer}
\end{figure}

To further probe the properties of the above-discussed two-component droplet phases (including the mixed ones), we numerically simulate the standard RF experimental protocol to dynamically create these droplet-gas mixtures~\cite{semeghini2018self,cabrera2018quantum}.
Specifically, the system is initiated in the ground state, $\Psi_0$, of the $\ket{1}$ hyperfine level (see also Sec.~\ref{sec:Setup}) with $N=2\times 10^5$ atoms 
utilizing the single component GPE without the LHY correction term.
Subsequently, at $t=0$, a specific portion of atoms is immediately transferred to the $\ket{2}$ hyperfine level, thus realizing a two-component attractively interacting mixture exhibiting population imbalance $N_1/N_2$, while simultaneously satisfying $N_1+N_2 = N$. 
The wave function of each component after the RF transfer therefore obeys $\Psi_i= \sqrt{N_i/N} \Psi_0$, with $i=1,2$.
The involved scattering lengths, $a_{ij}$, before and after the atom transfer correspond to particular magnetic fields realizing different $\delta a$ values of $^{39}$K. Specifically, we consider $B=56.5, ~ 56.665,~ 56.845~$G, associated with $a_{11}= 33.35~a_0$, $a_{11} = 33.03~a_0$, $a_{11} = 32.69~a_0$ initially and $\delta a =-5.2, -3.5, -1.4~a_0$ after the RF respectively.
Moreover, in order to sustain the potential development of the gas fraction in the majority component for  imbalanced settings, a 3D spherical trap of $\omega_r/(2\pi) = \omega_z/(2\pi) = 5 ~ \rm{Hz}$ is assumed 
unless stated otherwise.

\begin{figure}[tb]
\centering  
\includegraphics[width=1\linewidth]
{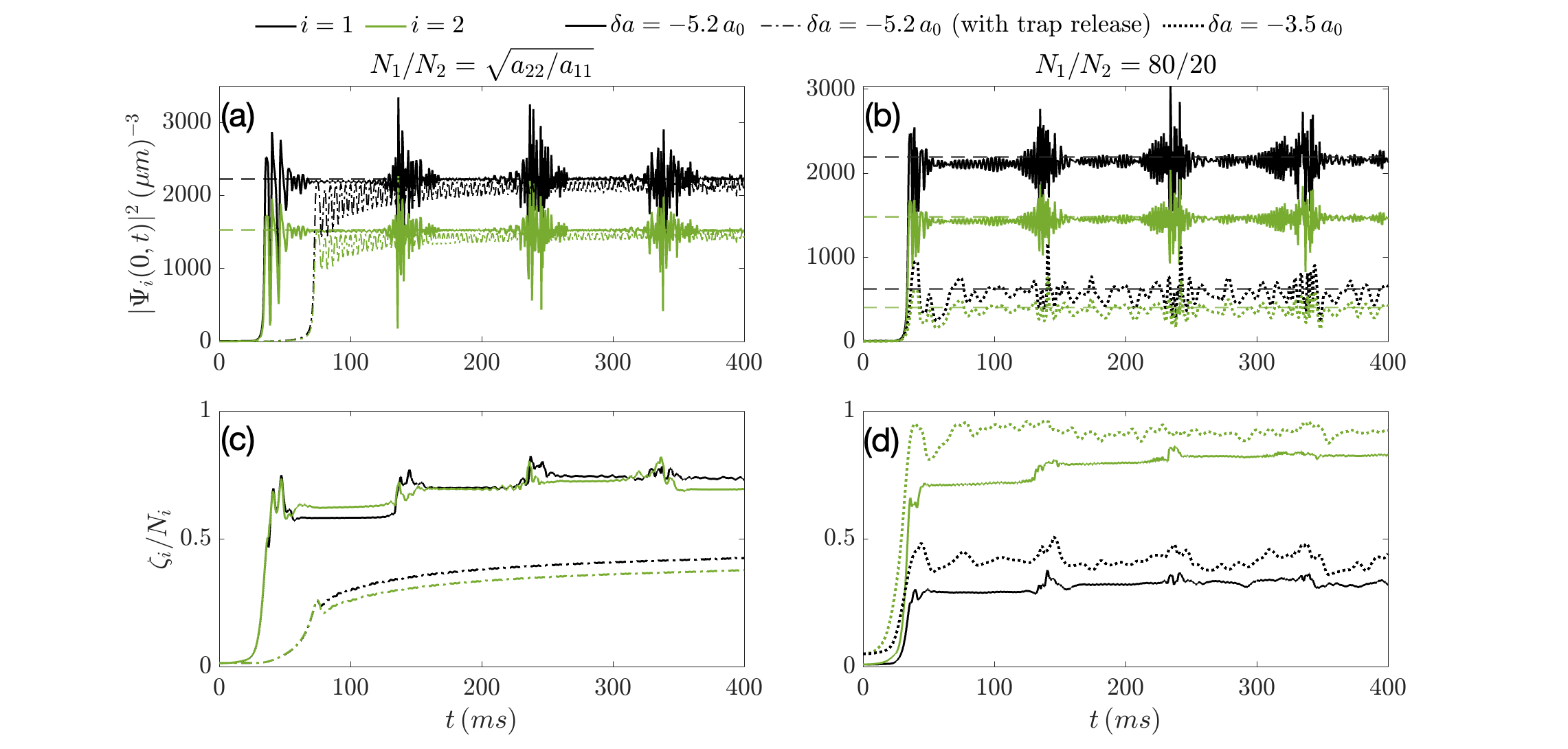}
\caption[]{Time-evolution of the peak densities $|\Psi(\boldsymbol{r}=0,t)|^2$ of both components (see legends) after the RF transfer in the case of (a) a balanced ($N_1/N_2 = \sqrt{a_{22}/a_{11}}$) and (b) an imbalanced ($N_1/N_2 = 80/20$) mixture for different mean-field attractions (see legends). 
The horizontal dashed lines represent the peak densities of the corresponding ground states characterized by the interactions after the RF protocol. The dash-dotted lines in panel (a) depict $|\Psi(\boldsymbol{r}=0,t)|^2$  associated with the RF transfer and a simultaneous trap release. 
(c), (d) Dynamics of the droplet fractions, $\zeta_i/N_i$ (see main text) for the RF used in panels (a) and (b) respectively.  
Even though the equilibrium densities are attained shortly after the droplet formation, the droplet fractions are significantly smaller than their ground state counterparts due to particle emission associated with self-evaporation.  
The latter causes the high amplitude oscillations of $|\Psi(\boldsymbol{r}=0,t)|^2$. 
}
\label{fig:RF_supplementary_data}
\end{figure}

We first focus on the particle balanced scenario characterized by $N_1/N_2 = \sqrt{a_{22}/a_{11}}$, at $\delta a = -5.2 ~ a_0$, so that we can initially understand 
the emergent population transfer dynamics in the simplest droplet setting  [Fig.~\ref{fig:RF_transfer}(a1), (b1)]. 
Due to the overall postquench mean-field attraction, the density of both components, $|\Psi_i(x,t)|^2 =   |\Psi_i(x,0,0,t)|^2$, contracts at short evolution times towards the trap center, giving rise to a high-density droplet at the same location around $t>40~$ms.
Comparing the dynamical behavior of the droplet peak densities [solid lines in Fig.~\ref{fig:RF_supplementary_data}(a)] with their respective ground state values [horizontal dashed lines in Fig.~\ref{fig:RF_supplementary_data}(a)], it turns out that both droplet cores attain their equilibrium densities rather quickly.
However, due to the significant difference between the initial (single-component) state before the RF transfer from the actual two-component droplet ground state, the dynamically created droplet cores become excited as the interactions are instantaneously quenched. As a consequence, they expel a jet of particles to lower their energy which is known as the self-evaporation process~\cite{petrov2015quantum,ferioli2020dynamical}.
This mechanism manifests by the large amplitude oscillations imprinted in   $|\Psi_i(0,t)|^2$ [Fig.~\ref{fig:RF_supplementary_data}(a)] 
caused by the particle emission that 
is clearly seen in Fig.~\ref{fig:RF_transfer}(a1), (b1). 
Due to the presence of the trap, the emitted atoms return back to the trap center further exciting the droplet core. 
As a result, the self-evaporation process occurs 
in a periodic manner.
The periodicity of this process depends crucially on the trap frequency, since the ejected atoms perform a breathing motion within the trap.

To better quantify the degree of proximity of the dynamically generated droplets to their ground state counterparts, we inspect the ensuing droplet fraction  ($\zeta_i/N_i$) by monitoring $\zeta_i = 4\pi \int_0^{R_d} dr ~ r^2 n_i(r) $. 
Here, $R_d$ is a radius associated with the ground state droplets as defined in Sec.~\ref{sec:core}.
As expected, following the droplet creation ($t>40~$ms) the fraction $\zeta_i/N_i$ increases rapidly [solid lines in Fig.~\ref{fig:RF_supplementary_data}(c)], but never reaches unity. 
This means that the size of the dynamically created droplets remains smaller than their respective ground state configurations [see also Fig.~\ref{fig:energy_N_per_da}(c)], as a result of the self-evaporation.
It is also noteworthy that as the ejected particles travel back towards the trap center, both droplet fractions slightly increase in an almost piecewise manner after every collision, attaining $\zeta_i/N_i \simeq 0.8$ at late evolution times.
Such collision events are accompanied by a small decrease of the energy of the droplet fragment towards more negative values (not shown for brevity), while during self-evaporation a portion of this energy is transferred to the expelled atoms. The latter have therefore sufficient energy to travel a bit further away from the trap center after consecutive collisions [see Fig.~\ref{fig:RF_transfer}(a1), (b1) at $t=200,~300,~400~$ms].

The distinction between the self-evaporation process and the trap effects outlined above can be achieved by emulating the RF transfer along with a simultaneous trap removal. 
Here, the released kinetic energy (caused by ramping down the trap) favors a dispersion of both clouds that is responsible for a deceleration of the inevitable density contraction. As such, the peak densities $|\Psi_i(0,t)|^2$ [dash-dotted lines in Fig.~\ref{fig:RF_supplementary_data}(a)] reach their equilibrium values [horizontal dashed lines] at somewhat later times compared to the trapped scenario.
The dynamically formed droplets are once more motionally excited, as evidenced by the oscillations of their peak densities [Fig.~\ref{fig:RF_supplementary_data}(a)].
However, the amplitude of these oscillations becomes gradually damped, hinting again to a self-evaporation process in order to lower the overall energy of the mixture.
Indeed, the relevant droplet fractions [dashed-dotted lines in Fig.~\ref{fig:RF_supplementary_data}(c)] reveal that $\zeta_i/N_i$ attain values much smaller than unity due to the particle emission.
In fact, the droplet segments are always smaller than their siblings in the trapped case, due to the expansion of the initial cloud after the trap release.

Next, we turn our attention to the nucleation of a particle imbalanced droplet with $N_1/N_2 = 80/20$ at $\delta a = -5.2 ~ a_0$, see Fig.~\ref{fig:RF_transfer}(a2), (b2).
After the RF protocol, the two droplet fragments emerge at the same time as in the balanced case [Fig.~\ref{fig:RF_transfer}(a1), (b1)] in both components, i.e. around $t \simeq 40~$ms, a behavior that is attributed to the same initial conditions in the two setups.
A significant difference with the balanced mixture is the rise of a gas fraction [see the orange shaded region in Fig.~\ref{fig:RF_transfer}(a2)] which surrounds the majority droplet segment.
This fraction persists throughout the evolution despite the disturbances caused by the expelled atoms of the majority droplet core traced back  to self-evaporation which is periodic due to the presence of the trap. 
Similarly to the balanced setting, such a process can be identified by the  
high amplitude oscillations of the droplet peak densities [solid lines in Fig.~\ref{fig:RF_supplementary_data}(b)].
Moreover, even though the equilibrium peak densities are attained shortly after the droplet formation [Fig.~\ref{fig:RF_supplementary_data}(b)], the spatial extent of the dynamically formed droplet segments is 
smaller than the one pertaining to the respective ground states.  
This is imprinted in the temporal evolution of $\zeta_i/N_i$ shown in Fig.~\ref{fig:RF_supplementary_data}(d) [solid lines] where values  
smaller from the equilibrium ones are eventually reached [Fig.~\ref{fig:energy_N_per_da}(c)] as a result of the self-evaporation.
In particular, the droplet segment of the majority component 
remains relatively low, namely $\zeta_1/N_1 \simeq 0.3$, reflecting the droplet-gas coexistence. 

To appreciate the impact of interactions in the dynamical formation of two-component imbalanced droplets, a RF protocol is followed to less attractive  averaged mean-field couplings  
$\delta a = -3.5 ~ a_0$.
This results in the formation of droplet segments in both components 
at slightly larger times, as can be inferred by the saturation of their peak densities [dotted lines in Fig.~\ref{fig:RF_supplementary_data}(b)].
Moreover, in contrast to $\delta a = -5.2 ~ a_0$, the droplet fractions acquire somewhat higher values [dotted lines in Fig.~\ref{fig:RF_supplementary_data}(d)], being closer to their equilibrium ones [see also Fig.~\ref{fig:energy_N_per_da}(c)].
This is a consequence of a less pronounced self-evaporation mechanism at $\delta a = -3.5 ~ a_0$ compared to $\delta a = -5.2 ~ a_0$, since in the former case the initial state has a larger overlap with the respective ground states~\cite{ferioli2020dynamical}. 
This fact can be clearly deduced from the finite droplet fractions $\zeta_i/N_i$ at $t=0$ in Fig.~\ref{fig:RF_supplementary_data}(d).
Finally, tuning $\delta a$ to even smaller attractions [Fig.~\ref{fig:RF_transfer}(a3), (b3)], e.g. $\delta a = -1.4 ~ a_0$, the RF transfer leads to a gaseous two-component state in line with the ground state predictions [Fig.~\ref{fig:1D_density_slices}(a3)] which naturally undergoes a breathing motion in both components due to the interaction quench. 
The period of the collective breathing dynamics is similar to the above-discussed revivals of the ejected atoms, related to self-evaporation that occurs for larger attractions, since this collective mode is again dictated by the external confinement. 
As expected, introducing three-body losses in the droplet-gas mixture dynamics leads to the dissociation of the RF generated central droplet structures within a few ms, see  Appendix~\ref{sec:three_body_loss} for more details.

\section{Conclusions and perspectives} \label{sec:Conclusions}

We have studied the  ground-state phases and quantum dynamics of 3D particle imbalanced two-component bosonic mixtures experiencing an average mean-field attraction. To compensate the attraction induced wave collapse, the first-order LHY quantum correction contribution is exploited for stabilizing the mixture into  self-bound configurations. 
Our investigation is based on the  eGPE framework within parameter regimes dictated by droplet experiments and explores the impact of isotropic and anisotropic 3D external confinement. 
For larger intercomponent particle imbalances and strong attractions, mixed droplet-gas phases occur.  
These refer to states where the minority component assembles in a droplet and the majority splits into a droplet fragment, whose atoms bind to the minority droplet, and the remaining particles lie outside of it in a gas segment. 
The atom number ratio between the aforementioned droplet fragments is adequately dictated by the density ratio locking condition. 

We showcase different ways to manipulate the portion of the gas segment. 
Specifically, it can be enhanced by employing i) larger atom numbers, ii) increasing intercomponent particle imbalanced ratios, iii) using anisotropic confinements,  
or iv) by utilizing box traps of smaller size. 
Accordingly, the binding energy of these mixed phases is larger not only for stronger attractions but also for smaller particle imbalances or decreasing trap aspect ratio attaining the isotropic 3D confinement. 
In fact, reducing the intercomponent particle balance gradually eliminates the gas segment resulting solely in droplet states in both components. 
Interestingly, a gradual transition from the mixed droplet-gas to a gas phase takes place for smaller in magnitude averaged mean-field attractions. 
To gain further insights into the mixed configurations we formulate an appropriate VA based on the modeling of droplet (gas) fragment by a super-Gaussian (Thomas-Fermi) profile. 
This is shown to be able to replicate both constituents for different parametric variations and actually estimate their behavior in sync with the eGPE predictions.

Next, the dynamical response of the different droplet-gas and gas phases is explored by simulating the routinely used experimental techniques of TOF expansion and RF transfer. 
In the former case, the two-component configurations are prepared within the external trap which is subsequently released letting the mixture  evolve freely. 
It is found that the participating droplet segments are retained robust in the course of the evolution, whilst the gas portion undergoes expansion. 
These features confirm the self-bound nature of the droplet fractions and the gas character of the excess atoms. 
The expansion rate of the gaseous atoms is faster for anisotropic traps due to the larger in-plane potential energy. 
Interestingly, we show that the expanding boundaries of the gas fragment follow the semi-analytical prediction (within the hydrodynamic prescription) for the expansion radius of the single-component quantum gas. 
This result verifies once more the gas character of the excess atoms. 
Turning to the RF atom transfer from one hyperfine state where the system assembles in a gas state to two individual particle-imbalanced ones suitable for droplet realization, we observe the dynamical formation of the different droplet-gas and gas states. 
Here, the droplet fragments form within a few milliseconds becoming excited and hence  emitting excitations due to self-evaporation. 
Simultaneously, a gas segment stemming from the original gas state forms and surrounds the droplet fragment of the majority component. 
Following either the TOF or RF processes we deduce that the number of bound state atoms in the individual components approximately satisfies the density ratio locking especially for smaller negative attractive interactions.  

There is an array of interesting extensions of our results which pave the way for probing quantum fluctuation phenomena in the nonequilibrium dynamics of many-body self-bound states. 
The effects of three-body losses in the real-time propagation of imbalanced droplets which we touch upon during the RF dynamics in Appendix~\ref{sec:three_body_loss} can be also considered in the TOF dynamics to facilitate  direct comparisons with experimental data. Moreover, the emulation of injection RF  spectroscopy~\cite{massignan2025polarons} by considering a spinor minority species including Rabi-coupling between the components and detuning would allow to explore the dynamical formation of droplet resonances and possibly identify excited (metastable) states. 
Here, the transition from quasi-particle (polaronic) to droplet states by tuning the population of the minority component is desirable.  
Along these lines, exploring and characterizing the emergent pattern formation, including shock- and rogue waves due to the imposed Riemann initial conditions featured by the mixed droplet  configurations due to their sharp boundaries, by dynamically crossing the distinct phases is worth pursuing. 
In addition, generalization of the two-component droplet framework within the mean-field stability regime as it was recently done in one-dimension~\cite{englezos2025multicomponent} shares the premise to reveal unseen exotic droplet configurations. 

 \begin{acknowledgments}

This work was supported by the Okinawa Institute of Science and Technology Graduate University. 
The authors are grateful for the Scientific Computing and Data Analysis (SCDA) section of
the Research Support Division at OIST. T.F. acknowledges support from JSPS KAKENHI Grant
No. JP23K03290. T.F. and T.B. are also supported by JST Grant No. JPMJPF2221. 
S.I.M acknowledges support from the Missouri University of Science and Technology, Department of Physics, Startup fund.

 \end{acknowledgments}

\appendix

\section{Impact of three-body losses in two-component droplets }\label{sec:three_body_loss}

\begin{figure}[tb]
\centering  
\includegraphics[width=1\linewidth]
{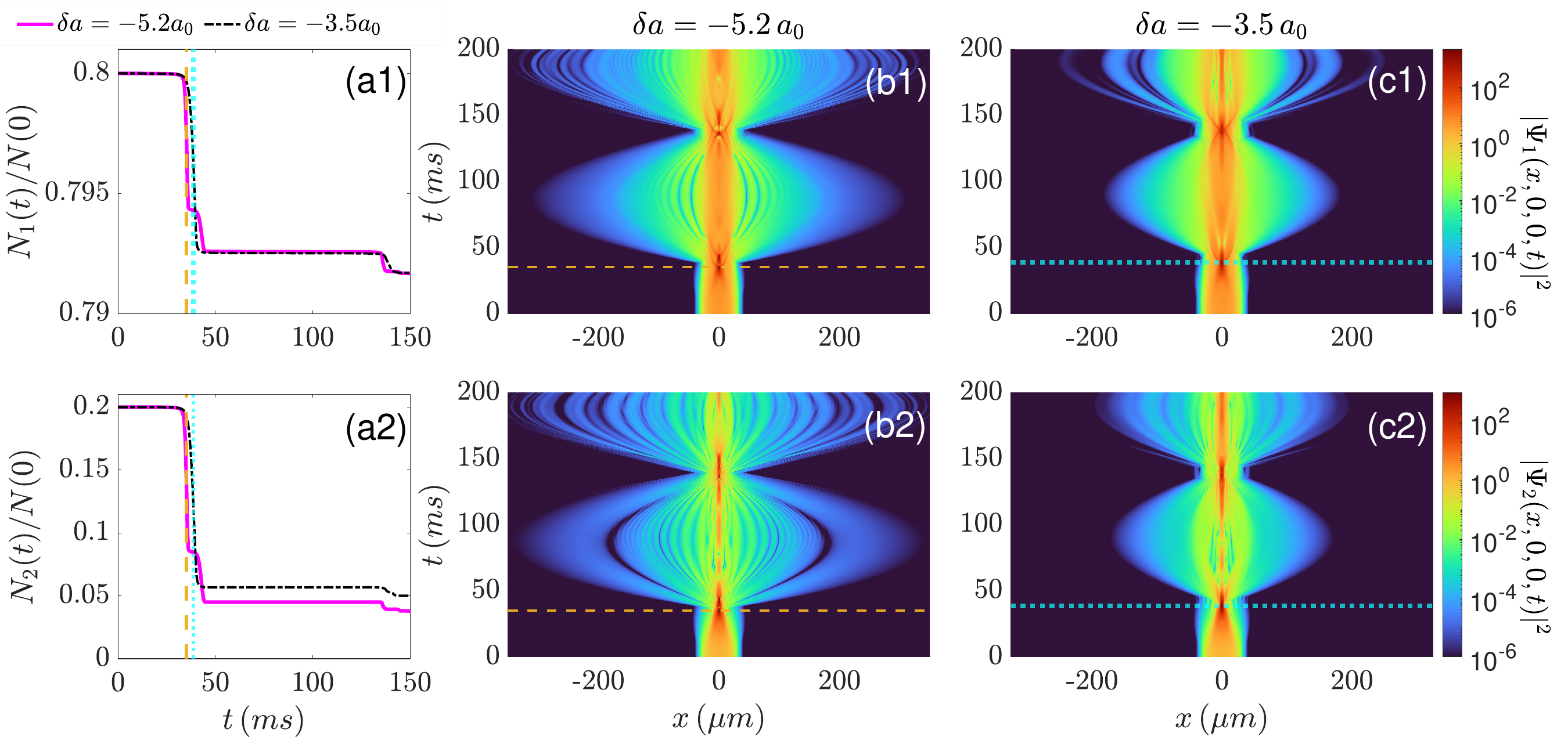}
\caption[]{
Dynamics of a $80/20$ particle imbalanced mixture generated by the RF transfer from a single-component gas, in the presence of three-body losses.
(a1) [(a2)] Evolution of normalized majority [minority] component particle number, $N_1(t)/N(0)$ ($N_2(t)/N(0)$), at different $\delta a$ (see legend).
(b1), (c1) [(b2), (c2)] Droplet-gas [droplet] generation in the majority [minority] component and subsequent decay due to three-body losses.
The dashed (dotted) lines in all panels mark the 
times at which the droplet-gas mixture dynamically transitions to a gaseous state  for different $\delta a$ (see legends). 
All other parameters are the same as in Fig.~\ref{fig:RF_transfer}.}

\label{fig:three_body_loss}
\end{figure}

Once formed, quantum droplets suffer from non-negligible three-body losses due to their high densities. 
In recent experiments, the lifetime of ${}^{39}$K droplets was measured to be around a few ms~\cite{semeghini2018self}.
Here, we examine the effect of such three-body losses on particle-imbalanced mixtures sustaining droplet-gas phases.
The latter are dynamically formed during the RF transfer of a single-component gas to an $80/20$ particle imbalanced configuration at various averaged scattering lengths [see also Sec.~\ref{rf_droplets} for further details].
In particular, three-body losses are taken into account via the phenomenological imaginary terms $-i \hbar K_{jjj} \abs{\psi_j}^4/2, ~ j=1,2$ in the coupled eGPEs [Eq.~\eqref{eq:eGPE}].
The three-body recombination rates are obtained from the experimental work of Ref.~\cite{semeghini2018self} and correspond to $K_{{\rm 111}}/3! = 1\times10^{-29}\,{\rm cm^6/s}$, and $K_{{\rm 222}}/3! = 9\times10^{-28}\,{\rm cm^6/s}$ for the majority and the minority components, respectively. 

At short evolution times, similarly to the RF dynamics without three-body losses, the atoms (especially of the minority component) tend predominantly towards the trap center due to the prominent averaged mean-field attraction. This behavior is subsequently counterbalanced by the gradual increase of the repulsive LHY contribution, resulting in a droplet-gas mixture at $t \simeq 35 ~ ms$.  
During this process, the peak densities of both components increase [see also Fig.\ref{fig:three_body_loss}(b1)-(c2)], approaching the corresponding densities of the stationary droplet-gas configurations [see  Fig.~\ref{fig:RF_supplementary_data}(a), (b)].
At this stage, three-body recombination becomes important, leading to the particle decay as can be seen in Fig.~\ref{fig:three_body_loss}(a1),(a2).
Naturally, since $K_{{\rm 222}}$ is much larger than $K_{{\rm 111}}$, the minority component particle number $N_2(t)$ decays faster than $N_1(t)$ [Fig.~\ref{fig:three_body_loss}(a1),(a2)].
This behavior can also be seen from the density evolution, $\abs{\Psi_i(x,0,0,t)}^2$, [Fig.~\ref{fig:three_body_loss}(b1)-(c2)] where $\abs{\Psi_2(x=0,0,0,t)}^2$ decreases rapidly after the formation of the droplet, whereas, the decrease in $\abs{\Psi_1(x=0,0,0,t)}^2$ is comparatively smaller.
Nevertheless, the substantial depletion due to three-body losses destroys the droplet cores in both components [cf. Fig.~\ref{fig:RF_transfer} and Fig.~\ref{fig:three_body_loss} close to the trap center].
At $\delta a = -5.2 ~ a_0$ [$-3.5 ~ a_0$] the droplet cores are degraded at around $37 ~ ms$ [$42 ~ ms$], with the gaseous transition being indicated by the dashed [dotted] lines in Fig.~\ref{fig:three_body_loss}.
Droplet-gas mixtures formed at small $\abs{\delta a}$ survive for longer evolution times compared to mixtures at larger $\abs{\delta a}$, due to their smaller flat-top density values.
The timescale of droplet core suppression is estimated from the time-evolution of the averaged mean-field energy contribution.
We have confirmed within  our simulations that when the droplet core is destroyed, the averaged mean-field energy turns positive, while simultaneously the LHY energy term becomes substantially suppressed. This signals the formation of a gaseous mixture. 
The averaged mean-field energy term remains positive even when a high density peak appears at $t\simeq 130 ~ ms$ [Fig.~\ref{fig:three_body_loss}(b1)-(c2)].
This peak is associated with the breathing motion of the gas after the droplet dissociates and it is traced back to the presence of the harmonic trap.
Since the density at the trap center becomes again relatively enhanced, three-body losses yield further depletion in both components.
This behavior is clearly imprinted in the step-like behavior of $N_i(t)/N_0$ at $t\simeq 130 ~ ms$ [Fig.~\ref{fig:three_body_loss}(a1), (a2)].

\section{Three-component description of a particle  imbalanced Bose-Bose  droplet}\label{sec:droplet_reconstruction}

As discussed in the main text, particle-imbalanced Bose-Bose mixtures can lead to droplet-gas coexistence  in the majority component at relatively strong  attractions. This exotic phase of matter along with its variational reconstruction  [see Sec.~\ref{Sec:Variational}], motivates also the development of a three-component setup for its description. Specifically, the aim of this three-component setting is to capture i) the droplet fragment, $\bar{\psi}_1$, and ii) the gas segment of the majority component, $\psi_g$, as well as iii) the droplet configuration, $\psi_2$, building upon the minority component. Here, $\psi_2$ interacts only with $\bar{\psi}_1$, an assumption that is justified by the vanishing spatial overlap between the gas fraction and the minority droplet component, see also Fig.~\ref{fig:GS_density}(a), (b). In addition, the LHY term is omitted for the description of $\psi_g$, since this contribution is negligible for small densities and it is commonly absent in the gas phase. 
Accordingly, the energy functional of such a three-component  system reads 
\begin{subequations}
\begin{align}
\mathcal{E} &= \int d^3\boldsymbol{r}\left[\mathcal{E}_{\text{kin}}+\mathcal{E}_{\text{trap}}+\frac{1}{2}g_{11} n_1^2 + \frac{1}{2}g_{22}n_2^2 + g_{12}\bar{n}_1n_2+\frac{8m^{3/2}}{15\pi^2\hbar^3}(g_{11}\bar{n}_1 + g_{22}n_2)^{5/2}\right], \label{eq:energy_func_3_component} \\
\mathcal{E}_{\text{kin}} &= \frac{\hbar^2}{2m}(\abs{\nabla\sqrt{n_1}}^2+\abs{\nabla\psi_2}^2), \label{eq:E_kin_3_component}\\
\mathcal{E}_{\text{trap}} &= \frac{m}{2}(\omega_r^2x^2+\omega_r^2y^2+\omega_z^2z^2)(\bar{n}_1+n_2+n_g),
\end{align}
\end{subequations}
where $n_1 = \bar{n}_1 +n_g$, containing $  \bar{n}_1 = \abs{\bar{\psi}_1}^2$, $ n_2 = \abs{\psi_2}^2$, and $  n_g  =  \abs{\psi_g}^2 $. 

The minority component is normalized as before, namely $N_2 = \int d^3\boldsymbol{r}~n_2$, while the distributions of the  droplet and gas fragments of the majority component satisfy $\bar{N}_1 = \int d^3\boldsymbol{r}~\bar{n}_1$ and $N_g = \int d^3\boldsymbol{r}~n_g$ respectively, with $N_1=\bar{N}_1+N_g$. Since both $\bar{N}_1$ and $N_g$ are not known \textit{a-priori}, the fraction $\bar{N}_1 / N_1$ is treated as a variational parameter in order to minimize the overall energy. 
In this context, apart from the assumptions regarding the interspecies interactions and the LHY term, the gas, $\psi_g$, and the droplet, $\bar{\psi}_1$, fragments are treated as subsystems of the majority component.
Hence, they share the same kinetic term [see Eq.~\eqref{eq:E_kin_3_component}], and intraspecies interaction contribution [see $\frac{1}{2} g_{11} n_1^2$ in Eq.~\eqref{eq:energy_func_3_component}].
The minimization of the above energy functional with respect to the three wavefunctions leads to the following coupled eGPE-GPE model 
\begin{subequations}
\begin{align}
i\hbar\frac{\partial}{\partial t}\bar{\psi}_1 &= \left(-\frac{\hbar^2}{2m}\frac{1}{\sqrt{n_1}}\nabla^2(\sqrt{n_1}) + V_{\text{trap}} + g_{11}n_1 + g_{12}n_2 + \frac{4m^{3/2}}{3\pi^2\hbar^3}g_{11}(g_{11}\bar{n}_1 + g_{22}n_2)^{3/2}\right)\bar{\psi}_1 \label{eq:3_comp_droplet} \\
i\hbar\frac{\partial}{\partial t}\psi_2 &= \left(-\frac{\hbar^2}{2m}\nabla^2 + V_{\text{trap}} + g_{12}\bar{n}_1 + g_{22}n_2 + \frac{4m^{3/2}}{3\pi^2\hbar^3}g_{22}(g_{11}\bar{n}_1 + g_{22}n_2)^{3/2}\right)\psi_2
\label{eq:3_comp_minority}  \\
i\hbar\frac{\partial}{\partial t}\psi_g &= \left(-\frac{\hbar^2}{2m}\frac{1}{\sqrt{n_1}}\nabla^2(\sqrt{n_1}) + V_{\text{trap}} + g_{11}n_1\right)\psi_g,
\label{eq:3_comp_gas} 
\end{align}
\end{subequations}
where $V_{\text{trap}} = \frac{m}{2} (\sum_{k=x,y,z} \omega_k^2 k^2)$.

\begin{figure}[tb]
\centering  
\includegraphics[width=1\linewidth]
{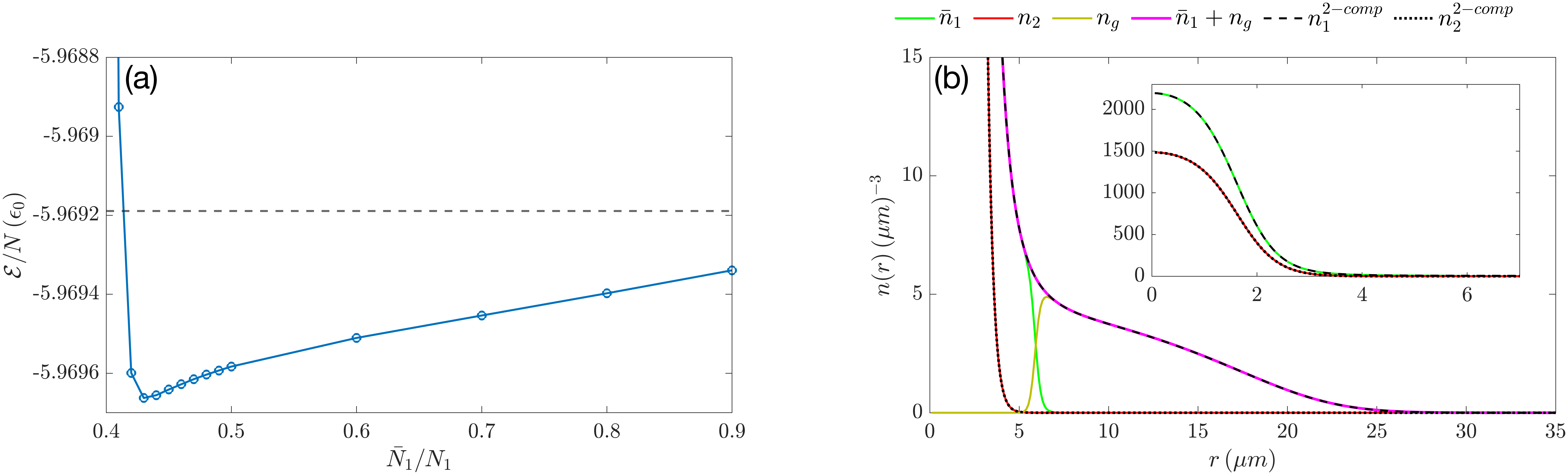}

\caption[]{(a) Energy per particle with respect to $\bar{N}_1/N_1$ and (b) density profiles of a three-component setup pertaining to a $80/20$ particle imbalanced Bose-Bose mixture at $\delta a =-5.2~a_0$ described by the coupled eGPE-GPE model of Eq.~(\ref{eq:3_comp_droplet})-(\ref{eq:3_comp_gas}). 
The energy minimum indicates the optimal fraction $\bar{N}_1/N_1$ for the lowest configuration.  
The dashed horizontal line in (a) marks the ground state energy per particle emanating from the original two-component eGPE description. A configuration with $\bar{N}_1/N_1=0.43$ yields very good agreement between the densities of the three-component system and the original bosonic mixture. The dashed and dotted lines in panel (b) refer to the densities obtained from the two-component eGPE description (see legend). 
}
\label{fig:3_component_data}
\end{figure}

The ground state energy of the three-component mixture is obtained by means of imaginary time evolution for different variational parameters $\bar{N}_1/N_1$. 
This process is presented in Fig.~\ref{fig:3_component_data}(a) for a $80/20$ imbalanced mixture at $\delta a = -5.2~a_0$ and trap characteristics $\omega_r=\omega_z = 2\pi \times 5~\rm{Hz}$.
The lowest energy is achieved for a configuration with $\bar{N}_1/N_1 = 0.43$, being very close to the droplet fraction predicted by the respective two-component eGPE [see also  Fig.~\ref{fig:energy_N_per_da}(c)].
More concretely, the relative energy error between the three-component setup and the original two-component eGPE description [horizontal dashed line in Fig.~\ref{fig:3_component_data}(a)] is of the order of $10^{-5}$, indicating a very good agreement between the two approaches. 
As a result, the density of the minority component stemming from the three-component setting agrees well with the respective ground state eGPE density [see the red solid and dotted lines in Fig.~\ref{fig:3_component_data}(b) and the  inset]. 
Moreover, the combination $\bar{n}_1 + n_g$ provides an adequate description of the majority component spanning both the droplet and gaseous regions. 
As a side note due to the repulsive interaction ($g_{11}$) the droplet and the gas segments of the majority component feature a small spatial overlap, similarly to the profile of a heteronuclear bosonic mixture with an impurity reported in Ref.~\cite{bighin2022impurity}.
As a result a relatively small tail of $\bar{n}_1$ exists behaving like a gas upon TOF expansion (not shown).

Summarizing, we can infer that the three-component model of the genuine particle-imbalanced droplet phase opens new avenues to study such mixed many-body configurations. A clear advantage is that, due to the separation between the gas and the droplet fragments, it is possible to monitor in a more clean way the propagation of excitations from the droplet to the gas fraction and vice versa. Furthermore, it facilitates the investigation of  three-body loss mechanisms in droplets, since the droplet segment can exchange particles with the gaseous fraction. Such a process can be important in prolonging the lifetime of these many-body self-bound states of matter and hence enable their experimental long-time observation.

\bibliographystyle{apsrev4-1}
\bibliography{references}

\end{document}